\documentclass{article} 

\usepackage[cmex10]{amsmath}

\usepackage{amsthm,amsfonts,amssymb}
\usepackage{times}
\usepackage{epsfig}
\usepackage{graphicx}
\usepackage{booktabs}
\usepackage{chngcntr}
\usepackage{authblk}
\usepackage[square,authoryear]{natbib}

\usepackage{pdfsync}

\include{macros3}

\usepackage[ruled]{algorithm2e}

\SetAlFnt{\small}
\SetAlCapFnt{\small}
\SetAlCapNameFnt{\small}
\SetAlCapHSkip{0pt}
\IncMargin{-\parindent}

\counterwithout{table}{section}

\begin{document}


\title{Audio Surveillance: a Systematic Review}

\author[1]{Marco Crocco - Marco.Crocco@iit.it}
\author[2]{Marco Cristani - marco.cristani@univr.it}
\author[3]{Andrea Trucco - trucco@dibe.unige.it}
\author[1,2]{Vittorio Murino - Vittorio.Murino@iit.it}
\affil[1]{Istituto Italiano di Tecnologia (IIT) - Pattern Analysis \& Computer Vision (PAVIS) Department}
\affil[2]{University of Verona - Dipartimento di Informatica}
\affil[3]{University of Genoa - Dipartimento di Ingegneria Navale, Elettrica, Elettronica e delle Telecomunicazioni (DITEN)}


\maketitle

\begin{abstract}
Despite surveillance systems are becoming increasingly ubiquitous in our living environment, automated surveillance, currently based on video sensory modality and machine intelligence, lacks most of the time the robustness and reliability required in several real applications. To tackle this issue, audio sensory devices have been taken into account, both alone or in combination with video, giving birth, in the last decade, to a considerable amount of research. In this paper audio-based automated surveillance methods are organized into a comprehensive survey: a general taxonomy, inspired by the more widespread video surveillance field, is proposed in order to systematically describe the methods covering background subtraction, event classification, object tracking and situation analysis. For each of these tasks, all the significant works are reviewed, detailing their pros and cons and the context for which they have been proposed. Moreover, a specific section is devoted to audio features, discussing their expressiveness and their employment in the above described tasks. Differently, from other surveys on audio processing and analysis, the present one is specifically targeted to automated surveillance, highlighting the target applications of each described methods and providing the reader tables and schemes useful to retrieve the most suited algorithms for a specific requirement.
\end{abstract}

\section{Introduction}

The monitoring of human activities has never been as ubiquitous and massive as today, with million of sensors deployed in almost every urban area, industrial facility and critical environment, increasing rapidly in terms of both amount and scope. In consequence of this, studies on automated surveillance have grown at a fast pace, with hundreds of algorithms embedded in various commercial systems.
In general, surveillance systems are based on one or more sensors able to acquire information from the surrounding environment. While the first generation of surveillance systems \citep{Raty2010} implied a monitoring activity by a human operator in order to detect anomalous situations or events, recently developed automated systems try to perform this task by computer vision and pattern recognition methodologies. The advantages of this perspective lie essentially in cost saving, especially with the decreasing price of sensors and processing units, and the ability to cope with huge amount of data (e.g. from tens or even hundreds of different sensors per surveillance system) which cannot be handled by human operators, not even for a short time.
\\


%

The early automated surveillance systems were based on one or more video cameras, such sensor typology being the most widespread also at the present days. Anyway, reasoning solely on visual data brings in considerable fallacies: among the others, the scarce performance of video cameras in adverse weather conditions, and their sensitivity to sudden light switching, reflections, shadows \citep{valera:192}. Moreover, standard video cameras are almost useless during nighttime, due to the scarce illumination and car flashlights.

To overcome these drawbacks, other kinds of sensors have been designed, exploiting them either alone or jointly with the video signal. In particular, near infrared or far infrared (thermal) cameras can substitute video cameras during nighttime or assist them, considerably improving the overall performance; in particular, thermal cameras are suited for the detection of hot objects against a colder background, such as people or moving vehicles \citep{Dai2005}. At the same time, infrared technology is highly dependent from the temperature, and the separation between background entities and foreground items can be problematic. \\
%

In this paper, we focus on the use of the audio information for surveillance purposes. This mean is less popular than the other modalities, especially in public surveillance systems, likely due to privacy issues, but it has been considered in many prototypical and research approaches \citep{pham2013streaming}.

Recording of audio stream provides a rich and informative alternative sensory modality both in indoor and outdoor environments. Among them, home interiors \citep{Zieger2009,Vacher2004}, offices \citep{Harma2005,Atrey2006}, banks \citep{Kotus2013}, elevators \citep{Radhakrishnan2005_1,Chua2014}, public transport vehicles \citep{Pham2010,Rouas2006,Vu2006}, railway stations \citep{Zajdel2007}, public squares \citep{Valensize2007}, farms \citep{Chung2013} or car parkings can by cited as particularly relevant for surveillance tasks, where audio can contribute consistently.

With respect to video sensors, audio sensors (as microphones) bear several appealing features:
\begin{itemize}
\item Audio stream is generally much less onerous than video stream and this fact encourages both the deployment of an higher number of audio sensors (also thanks to a lower unitary cost) and a more complex signal processing stage
\item While standard cameras have a limited angular field of view, microphones can be omnidirectional, i.e., with a spherical field of view
\item Due to the bigger involved wavelenght, many surfaces allow specular reflections of the acoustic wave,  permitting to acquire audio events also when obstacles are present along the direct path (even if this fact can be a drawback for sound localization task) 
\item Illumination and temperature are not issues for the audio processing
\item Several audio events important for surveillance task like shouts or gunshots have little or no video counterpart
\item From the psychological point of view, audio monitoring is experienced as less invasive than video monitoring, and can be a valid substitute in all situations in which privacy concerns are stressed \citep{Chen2005}. To this end, it is important to remark that audio surveillance does not usually include automatic speech detection and recognition.
\end{itemize}

\begin{figure}[t]
\centering
\includegraphics[width=8cm]{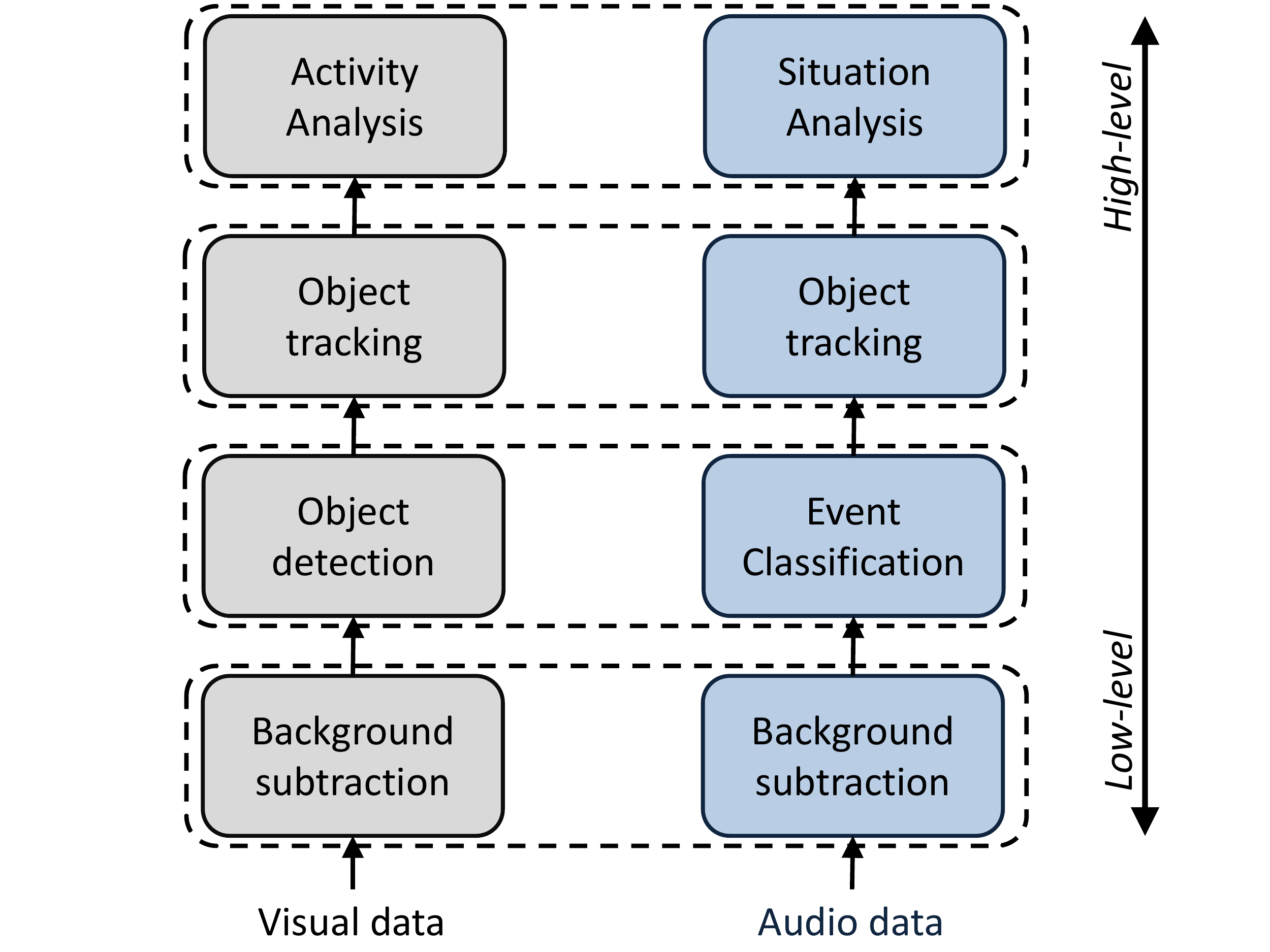}
\caption{Parallelism between the standard video surveillance workflow (on the left) and the audio surveillance pipeline (on the right).}
\label{fig:mainscheme}
\end{figure}

Despite automated audio surveillance is at its early steps, in the last decade a consistent amount of works have been published, and this paper contributes in providing the first systematic review. Several different taxonomies can be proposed to organize all the approaches, based either on the algorithmic nature of a method or its particular applicative scenario. In this paper, we organize the review by considering the different tasks where audio information can be exploited, in a typical surveillance framework. Borrowing from the more established and widespread video surveillance literature, which organizes the processing starting from low-level to high-level processing, four typical tasks have been identified: \emph{background subtraction,
event classification, object tracking and situation analysis}, see Fig.~\ref
{fig:mainscheme}. \\

Background subtraction is usually the first stage in a surveillance system, which aims at filtering the raw data, pruning away useless information, and separating expected data (the background) for interesting/unexpected items (the foreground). In video surveillance, background subtraction usually highlights moving objects, suppressing the visual content of the scene \citep{Cristani2010}.
The goal of audio background subtraction is analogous: to discard the expected audio signal, highlighting interesting audio events, that can be successively modeled and categorized \citep{CristaniICPR2004}.

While in video surveillance the background is often mostly static or slowly varying, the audio background exhibits an higher degree of variability due to the intrinsic time-varying nature of many sounds. Moreover, the audio signal is more complex due to the superimposition of multiple audio sources and to the multi-path propagation resulting in echo and reverberation effects. Whereas in the video analysis the background subtraction is limited to marking some pixels values, producing foreground masks, here the task is much more complex, incorporating source separation and filtering issues. Furthermore, the signal to noise ratio (SNR) is typically lower in an audio signal than in a video one, especially if the microphone is not very close to the acoustic source. All these issues make the audio background subtraction problem a challenging task. \\

Once the foreground is detected, the second stage in a surveillance system consists in characterizing the atomic entities lying therein.  In the video analysis, this operation consists in the classification of objects of interests into a set of predefined categories, like pedestrians, vehicles, animals and so on. This happens usually by employing heterogeneous set of features, fed into statistical classifiers. In the audio realm, basic entities of interests are called \emph{events}, defined by temporal windows and characterized by a particular spectral blueprint. Apart from the surveillance context, classification and separation of audio signals in a given environment is mainly labeled as Computational Auditory Signal Analisys (CASA) \citep{Bregman2005}. \\
%
%
%

The third typical task in video surveillance is localization and tracking of a moving object or person in a scene, producing spatial trajectories that can be employed afterwards for analyzing structured activities. In the audio context, the spatial tracking can be carried out only if spatialization is performed: actually, the use of multiple microphones placed in different locations allows to spatially sample the audio wavefield so recovering spatial information about the direction of arrival of an audio wavefront \citep{Choi2005}, the location of an audio source \citep{Huang2001} or even an acoustic map of the environment. In the latter case, the microphone array is used as an acoustic camera in order to obtain two or even three dimensional images of the acoustic intensity \citep{O'Donovan2007}.

In this way, the tracking problem can be decomposed in a sequence of spatial localization tasks, in which a particular sound source is localized over a short temporal window. Alternatively, localization can be considered as an input for a standard tracking algorithm, like Kalman filter or Particle filter, which relies on an underlying model of the source location dynamic and measurement noise. Multipath propagation and low SNR make the localization and tracking problem harder than the related video counterpart and the spatial resolution usually achieved can not reach the video one.\\

Finally, once multiple significant sounds are detected, classified and possibly tracked, all the information can be used together and combined into a higher analysis stage, in order to understand the nature of the scenario monitored. In video surveillance, this step is characterized by activity analysis: once objects have been characterized and tracked, this step provides a global characterization of what is happening in the monitored scene.\\

In the audio counterpart, this step is strictly linked to the so called Computational Auditory Scene Recognition (CASR) \citep{Peltonen2002}, \citep{Cowling2003}, aimed at the overall interpretation of an acoustic scene rather than the analysis of single sounds: this task is usually known as situation analysis. The situation analysis, due to its inherent complexity, has been addressed by relatively few approaches in literature, but it represents the final goal for an automatic surveillance system able to extract semantic information from the monitored scene.\\

The choice of a proper set of audio features is a crucial step, affecting all the four tasks above described. The complex nature of audio signal has encouraged the use of more sophisticated features in comparison to the video case. This led to a proliferation of cues mostly targeted to specific sounds or acoustic environments, but up to now it is not clearly established what set of features is the most performing in a general case. This fact is due partly to the lack of audio datasets taken as benchmarks by the audio community on which features can be tested, and partly to the fragmentation of the topic in the scientific literature among different fields as acoustic signal processing, pattern recognition, multimedia, etc.\\

As the information brought by audio sensors is in large measure complementary to the other modalities, a multimodal surveillance system can provide a more accurate and reliable performance. In this paper, in addition to audio-only surveillance tasks, we will also address multimodal, mostly audio-video, surveillance applications.

Multimodal tasks will be presented exploiting the taxonomy above introduced, since crossmodal fusion can also happen at different processing levels, low and high.
In addition, different sensory modalities can be generally fused at three different stages \citep{Atrey2010}: raw data-level, feature-level or decision-level. Our taxonomy will take into account this categorization too, in order to give the reader a precise snapshot of how the audio information interact with the other modalities. Nevertheless, it is worth noting that in the audio and video fusion, the raw data-level fusion is very rarely addressed due to the extreme difference in the two signals properties,
 while the other two hold sistematically.

The present paper is structured as follows: after the Introduction, Sections from 2 to 5 describe the methods devoted to background subtraction (Section 2), audio event classification (Section 3), audio source localization and tracking (Section 4) and  situation analysis (Section 5). In Section 6 audio features, often common to the above tasks, are classified in a general taxonomy and described in detail. Finally, in Section 7 some conclusions are drawn.


\section{Background subtraction}
Analogously to video \citep{Cristani2010}, audio background can identified by the recurring and persistent audio feature that dominates the portion of signal. Foreground sounds can be defined by the departure from the typical background features.

\begin{figure}[hb]
\centering
\includegraphics[width=8cm]{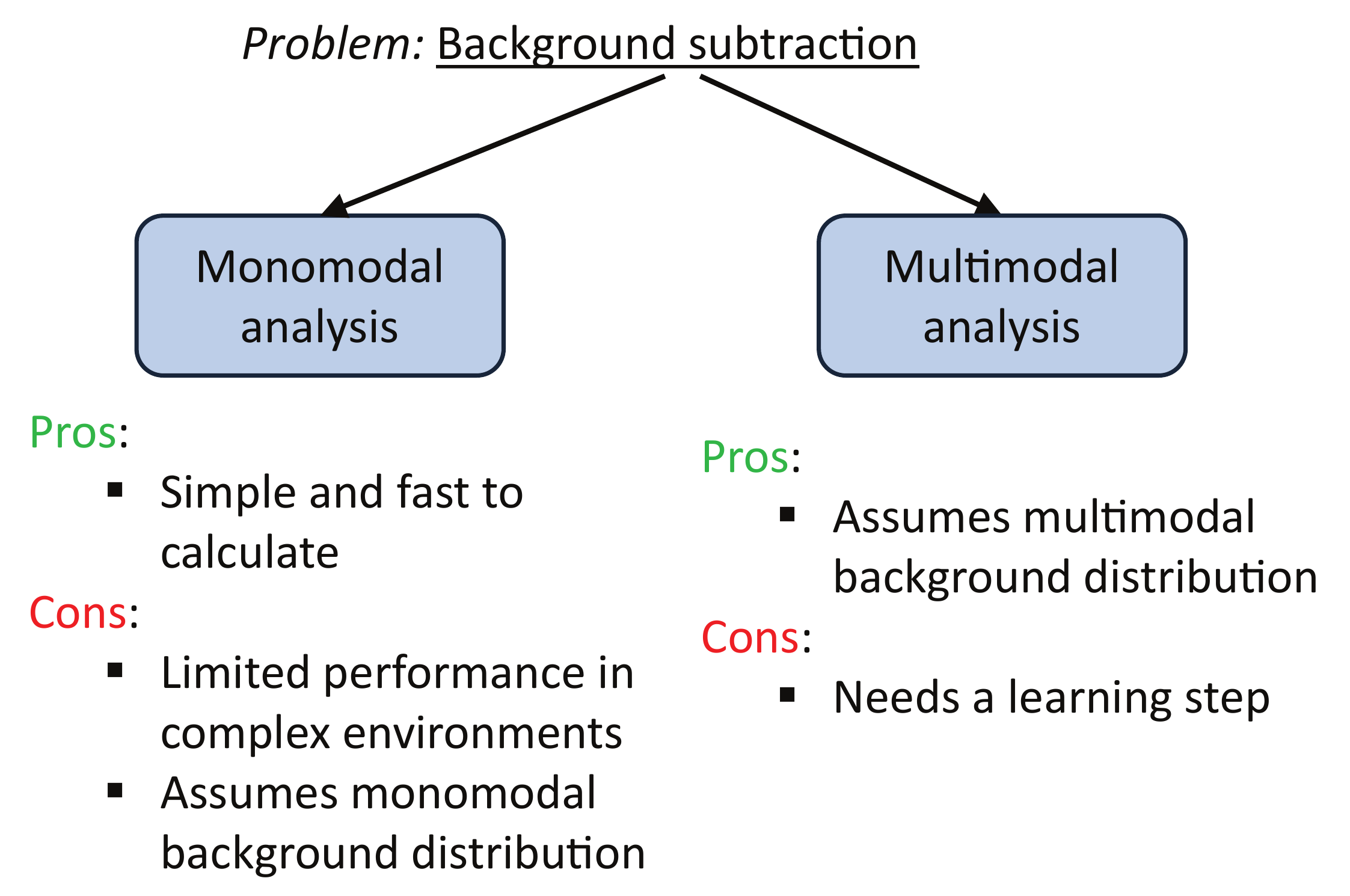}
\caption{General taxonomy of the audio background subtraction methods, with pros and cons added for each category of approach.}
\label{fig:BG_general}
\end{figure}

The background subtraction approaches can be divided into those techniques which operate a simple thresholding on the energy signal, implying that the distribution of audio features assumed as background is monomodal, and those approaches which perform a multimodal analysis, assuming that the audio background could be formed by different audio patterns that are repeated over time (see Fig.~\ref{fig:BG_general}).

\subsection{Background subtraction by monomodal analysis}
The most simple and intuitive parameter that can be used to discriminate a foreground sound is the signal energy.
Many sounds of interest, especially the impulsive ones like gunshots, door slams, cries cause an abrupt change in audio volume with respect to the typical auditory scenarios.
%
Following this principle, some works proposed to segment the audio stream into fixed-length windows and discard all the windows whose energy is below a predefined threshold \citep{Azlan2005} \citep{Smeaton2005}. Obviously, a criterion is needed to fix such threshold. The simplest way consists in analyzing a long portion of audio stream containing only background and fixing the threshold so as to capture louder sound signals \citep{Azlan2005}, or, alternatively, fixing the threshold proportionally to the average energy level \citep{Smeaton2005}. 

If the background average energy is known to vary deterministically, e.g. during nightime or daytime, the threshold can be tuned accordingly \citep{Smeaton2005}.

\begin{figure}[ht]
\centering
\includegraphics[width=8cm]{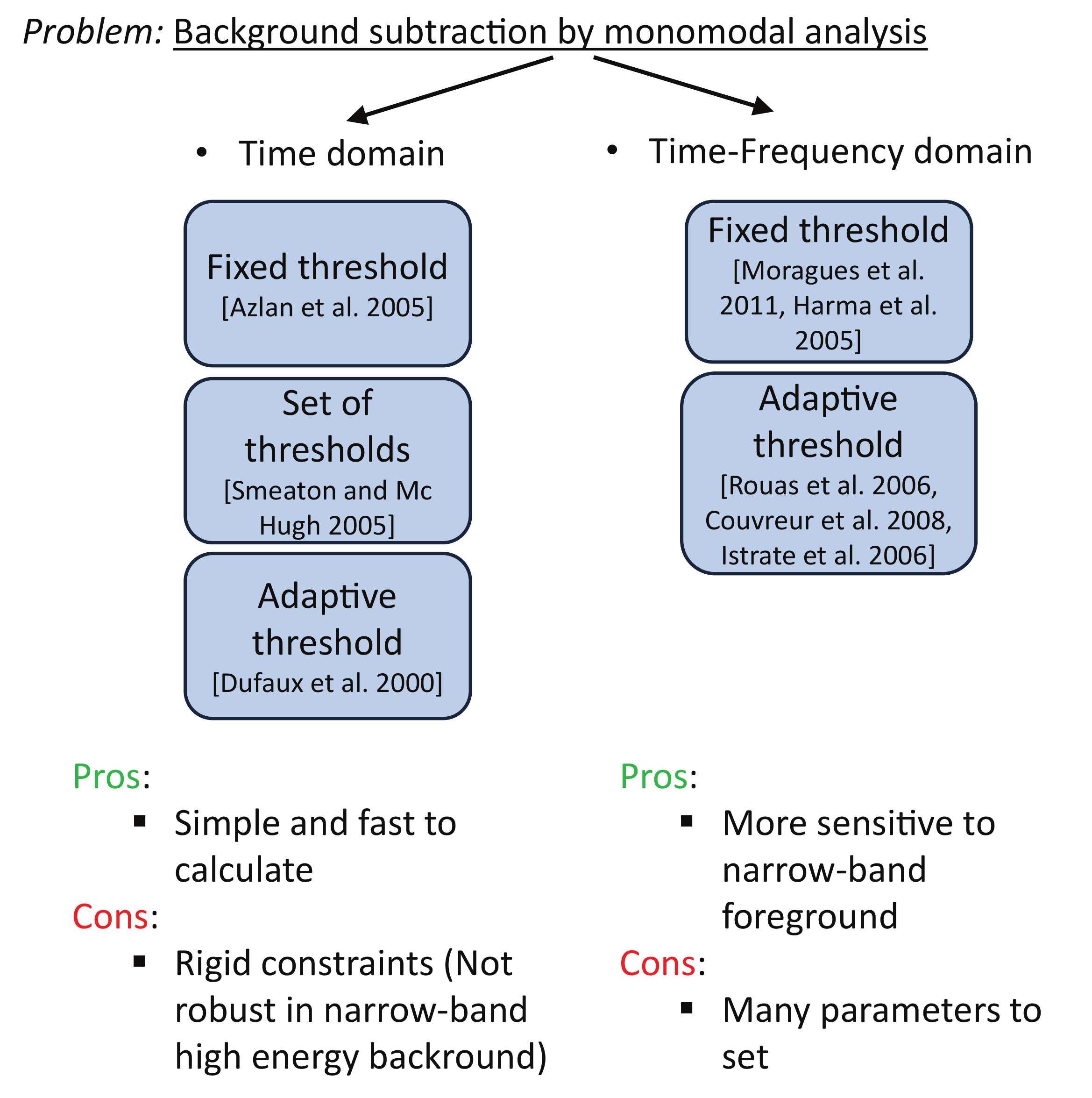}
\caption{Audio background subtraction by energy thresholding, with pros and cons added for each category of approach.}
\label{fig:BGthresh2}
\end{figure}

On the contrary, if the background energy variation in time is not predictable it is necessary to adopt an adaptive threshold. In \citep{Dufaux2000},  the signal energy, estimated over a number of temporal windows, is median-filtered, and the output of the filter is subtracted from the energy. The result is normalised, emphasizing the relevant energy pulses. Finally an adaptive thresholding, depending on the standard deviation of a past long-term windowed energy sequence, is applied. A scheme summarizing the BG subtraction by thresholding is reported in Fig.~\ref{fig:BGthresh2}.\\

To cope with foreground signals of variable duration and bandwidth, in \citep{Moragues2011} energy thresholding is applied in the time-frequency domain, that is, in parallel at different time scales and on different frequency bands. A foreground signal is considered to be detected if in at least one of the scale-frequency bins the energy is higher than the threshold.

The previous methods adopt a signal segmentation in block of fixed length: this fact may chop a significant audio event into two adjacent blocks so making more difficult the subsequent processing stages. To avoid this drawback, in \citep{Rouas2006} an Autoregressive Gaussian model is employed to predict the current audio sample on the base of the previous ones: if the prediction error is higher than a certain value it is assumed that around that sample a different sound arose and the temporal window boundary is fixed. Subsequently, each temporal window is classified into background or foreground on the base of an adaptive threshold.

The segmentation problem is extensively treated in \citep{Kemp2000}, where three different approaches are compared: energy-based, model-based and metric-based segmentation. Energy-based segmentation, the less performing one, puts simply a boundary every time a silence period between two audio events is detected.

Model-based segmentation trains a statistical model for every predefined class of audio sound and puts a boundary every time adjacent audio segments are classified into different classes.

Metric-based segmentation evaluates the distance between two adjacent segments by means of a pseudo-metric like Kullback-Leibler divergence,
putting a boundary if the distance is higher than a threshold. Interestingly, the threshold can be determined by an information theoretical measure like Minimum Description Length or Bayesian Information Criterion.

Finally, in \citep{Kemp2000} a hybrid strategy, which outperforms the previous methods is proposed: first a metric based clustering of short segments is performed; secondly, each cluster is considered as an audio class, whose segments are employed to train a statistical model, which is subsequently used to segment the audio stream.\\

Energy thresholding yields in general limited performance in complex environments where high energy sounds may periodically appear yet being part of the background, e.g. car engines in a car parking. In these cases, to improve the background subtraction it is useful to extract other features from the signal, besides the energy, and examining the departure of their values from the typical ones in order to detect foreground events. A comprehensive exposition of acoustic features employed in audio surveillance can be found in Section 6. Differently, in Table \ref{table:Feat_BG} features employed in background subtraction, classified according to the taxonomy described in Section 6, and the corresponding references are reported. Similar tables are reported in subsequent sections for the other tasks.

In \citep{Istrate2006}, wavelet coefficients are extracted from the signal and the energy of the upper coefficients, corresponding to the higher frequencies, is compared with an adaptive threshold. The rationale is that background noise has mostly low frequency components whereas foreground tends to be more impulsive.

 In \citep{Harma2005}, the differences between the frequency bins of the Fourier transform of the current window and the mean Fourier Transform are calculated. Then, in order to improve the detection of narrow band audio events, the difference between the maximum peak and the variance of the incremental frequency bins is evaluated and compared to a threshold.

In \citep{Couvreur2008}, the selected features are mainly drawn from psycho-acoustical findings on the human auditory attention system; three alternative normalization methods are proposed and a final threshold is adaptively determined by minimizing the sum of the intra-class variance over a given temporal window, where the two classes represents background and foreground sounds.

\subsection{Background subtraction by multimodal analysis}

In highly complex audio environments, the assumption that features values related to background are spread around a single value, i.e. that background feature values can be modeled as a unimodal distribution, may not hold anymore. In such a case, background modeling with multimodal distributions may provide better results in term of background/foreground discrimination.

A widely used multimodal model is the so called Gaussian Mixture Model (GMM), 
Given a vector of feature extracted from the signal their joint probability density function can be modeled as a sum of multidimensional gaussian functions with different mean vectors and covariance matrices.
The underlying idea is that each sound source corresponds to a Gaussian distribution of the mixture. Usually, in order to cope with an adaptive background, the mixture parameters are updated at each iteration using the current feature vector. Different criteria have been proposed to both update the mixture parameters and to discriminate between background and foreground given the mixture model.

In \citep{Radhakrishnan2005}, the GMM is trained over a background audio sequence using the Minimum Description Length method; subsequently, the probability of the current feature vector conditioned to the background GMM is calculated and compared to a predefined probability threshold. If the probability is lower than this threshold the observation is judged to be generated from a different probability distribution and so classified as foreground. Instead, if the current observation is classified as background, the GMM is updated by building a second GMM on the base of the most recent observations and fusing together the two GMMs by pairing and merging the most similar mixture components.

In \citep{Cristani2004}, the most likely mixture component that matches the current observation is found, choosing among a set of Gaussian components which are ranked in descending order with respect to their weight and divided by their variance. If the sum of the weights until the matched component is higher than a threshold the observation is labeled as foreground. If no match is found, as the observation is too far from every component, a new component is created substituting the one with the lowest weight. The parameters of the matched component, and all the weights are then updated, irrespective of the BG/FG classification. Differently from \citep{Radhakrishnan2005}, the GMM in \citep{Cristani2004} models explicitly both the background and the foreground. Another difference is that in \citep{Cristani2004} a uni-dimensional GMM is used for each feature (in that case the energy in a given frequency band) and the background/foreground classification is carried out independently for each feature. This choice allows a computational advantage but assumes the features to be independent from each other, which is typically a strict assumption.\\

 In \citep{Moncrieff2007}, a multidimensional GMM is employed analogously to \citep{Radhakrishnan2005}, but some solutions are proposed to deal with quite complex background environments. First, fragmented background states are unified by means of an entropy-based approach in order to avoid erroneous foreground classifications. Then the number of states is adaptively tuned according to the background complexity, and finally a cache is introduced to retain in memory the background components related to rapidly changing sounds.\\
In \citep{Ito2009}, the problem of rarely occurring background events is faced by means of a multi-stage GMM approach. In the training phase, the first stage GMM is trained over all the background samples available: the samples with resulting lower likelihood are used to train the second stage GMM and so on. In the testing phase a sample is definitely classified as foreground only if it is classified as foreground in each of the GMM stages.\\

\begin{figure}[ht]
\centering
\includegraphics[width=10cm]{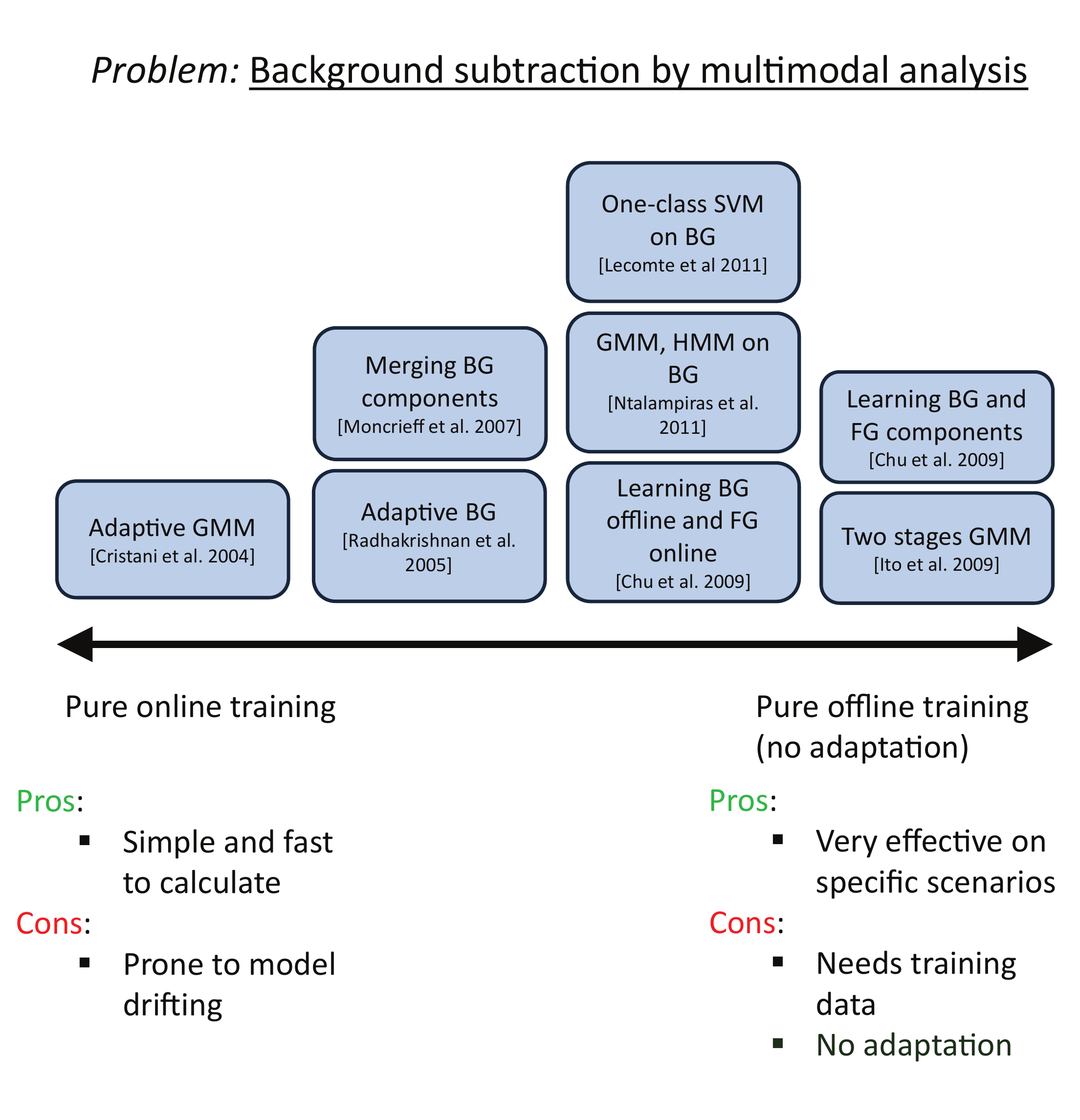}
\caption{Multimodal background subtraction by learning: the figures highlights those methods requiring more or less offline learning.}
\label{fig:BGmulti}
\end{figure}

A problem not addressed by the previous methods is the case of slowly varying and gradual foreground, like a plane passing overhead. In such a case adaptive methods tends to classify foreground as background because it is holding for a while. To overcome this drawback, in \citep{Chu2009_2} a semi-supervised method is adopted. First, both background and foreground are trained offline exploiting previous knowledge of specific foreground sounds; secondly, a separate model detects the changes in the background and it is finally integrated with the offline audio prediction models, that act as prior probabilities, to decide on the final foreground/background discrimination.\\

A different semi-supervised model, which specifically addresses the problem of detecting rare and unexpected foreground audio events, is proposed in \citep{Zhang2005}. Usual events, i.e. the background, are used to train offline an Hidden Markov Model (HMM); unusual event are learned online iteratively adapting the usual event model to unusual events by means of Bayesian Adaptation Technique. In this way it is possible to supply to the scarcity of unusual events in the offline training phase.\\

\begin{table*}[!hb]
\centering{
  { \scriptsize
\begin{tabular}{| p{1.5 cm} | p{5 cm} | p{6 cm}|}
\hline
 \textbf{Class} & \textbf{Short Description} & \textbf{Reference}  \\
\hline

 Time   &  Zero-crossing rate  & \citep{Couvreur2008,Moncrieff2007}.\\
            &  Waveform minimum and maximum & \citep{Ntalampiras2011}.\\
            &  Autocorrelation coefficients & \citep{Couvreur2008}.\\
\hline
 Frequency            &  Fourier coefficients & \citep{Harma2005}.\\
                      &  Fundamental frequency & \citep{Ntalampiras2011}.\\
                      &  Spectral flatness   & \citep{Ntalampiras2011}.\\
\hline
 Cepstrum             &  MFCC  &\citep{Radhakrishnan2005,Moncrieff2007,Ito2009,Ntalampiras2011,Chu2009_2,Zhang2005}.\\
                      &  MFCC derivatives  & \citep{Moncrieff2007}.\\
\hline
 Time-frequency       &  Wavelet coefficients & \citep{Istrate2006,Moncrieff2007}.\\
                      &  Mean and Std of frequency and scale of Gabor atoms selected by Matching Pursuit & \citep{Chu2009_2}.\\
\hline
 Energy               &  Signal energy over a fixed window & \citep{Azlan2005,Smeaton2005,Dufaux2000,Kemp2000,Cristani2004,Rouas2006,Ito2009,Moncrieff2007}.\\
                      &  Interaural Level Difference & \citep{Hang2010}, \citep{Hu2010}.\\
\hline
 Biologically or perceptually driven  &   Spectral features based on Gammatone filter bank: spectral moments, slope, decrease, roll-off, variation, flatness etc. & \citep{Couvreur2008}.\\
                      &  Mel frequency coefficients & \citep{Kemp2000}.\\
                      &  Intonation and Teager Energy Operator (TEO) based features & \citep{Ntalampiras2011}.\\
                      & High-order Local Auto-Correlation (HLAC) & \citep{Sasou2011}.\\
\hline
\end{tabular}
}
}
\caption{  Features employed in background subtraction: first column indicates the feature class according to the taxonomy defined in Section~\ref{sec:audiofeatures}, second column reports feature names and third column reports references of works where they are employed.}
\label{table:Feat_BG}
\end{table*}

A more challenging situation is faced in \citep{Ntalampiras2011},  where the class of anomalous events is assumed to be unbounded and only normal events, i.e. the background, are available in the training phase. Under the hypothesis that anomalous events are significantly different from the normal ones, this approach does not model explicitly the formers but labels as abnormal each event whose likelihood is lower than the lowest likelihood of the normal events belonging to the training set. Modeling of the normal event class is performed by means of three methods: GMM, HMM and GMM clustering. In the latter, several GMMs, each one corresponding to an audio recording of normal events, are modeled. A matrix of distances between the GMMs is computed using Kullback-Leibler similarity measure, and the GMM with the minimum distance from all the others is chosen as representative of the whole class of normal events. This approach showed better performance in case of complex environments with many different normal sounds in respect to standard GMM and HMM.

The same scenario of \citep{Ntalampiras2011} is faced in  \citep{Lecomte2011}, where One-Class Support Vector Machine (SVM) with Radial Basis Function (RBF) kernel is adopted to model the background sound scene and to detect the onset of anomalous events. One-Class SVM builds an hyperplane separating the feature space into background and foreground regions. Since only background is available in the learning phase, the optimization criterion consists in the trade-off between the empirical classification error on the background class and the volume of the feature space corresponding to the background. The smaller the volume the simpler the background model and the lower the structural risk of false background classifications. As GMM before, One-Class-SVM with RBF models the PDF over the feature space related to background class as a mixture of Gaussian functions, but has better generalization properties thanks to the volume penalty, automatically learns from data the proper number of mixtures and does not suffer from local minima in the learning phase. However, if online learning is required One-Class-SVM is less suited than GMM, mainly due to the higher computational load.\\

See Fig.~\ref{fig:BGmulti} for a graphical summarization of the models assuming the BG is multimodal.

Besides energy changes, audio events deserving attention are often characterized by a rapid movement of the sound source location, whereas audio background has a more static spatial characterization. Based on this principle, in \citep{Hang2010} sound source location is estimated by looking at interaural Level Difference (ILD) between a couple of microphones, followed by sound source velocity is estimation as the difference of ILDs between subsequent audio frames ($\Delta$ILD). To cope with multiple objects moving in different directions the $\Delta$ILD  is evaluated at several frequency bands and a final threshold is set multiplying the $\Delta$ILD mean by the $\Delta$ILD variance.

\section{Audio events classification }
\begin{figure}[ht]
\centering
\includegraphics[width = 15cm]{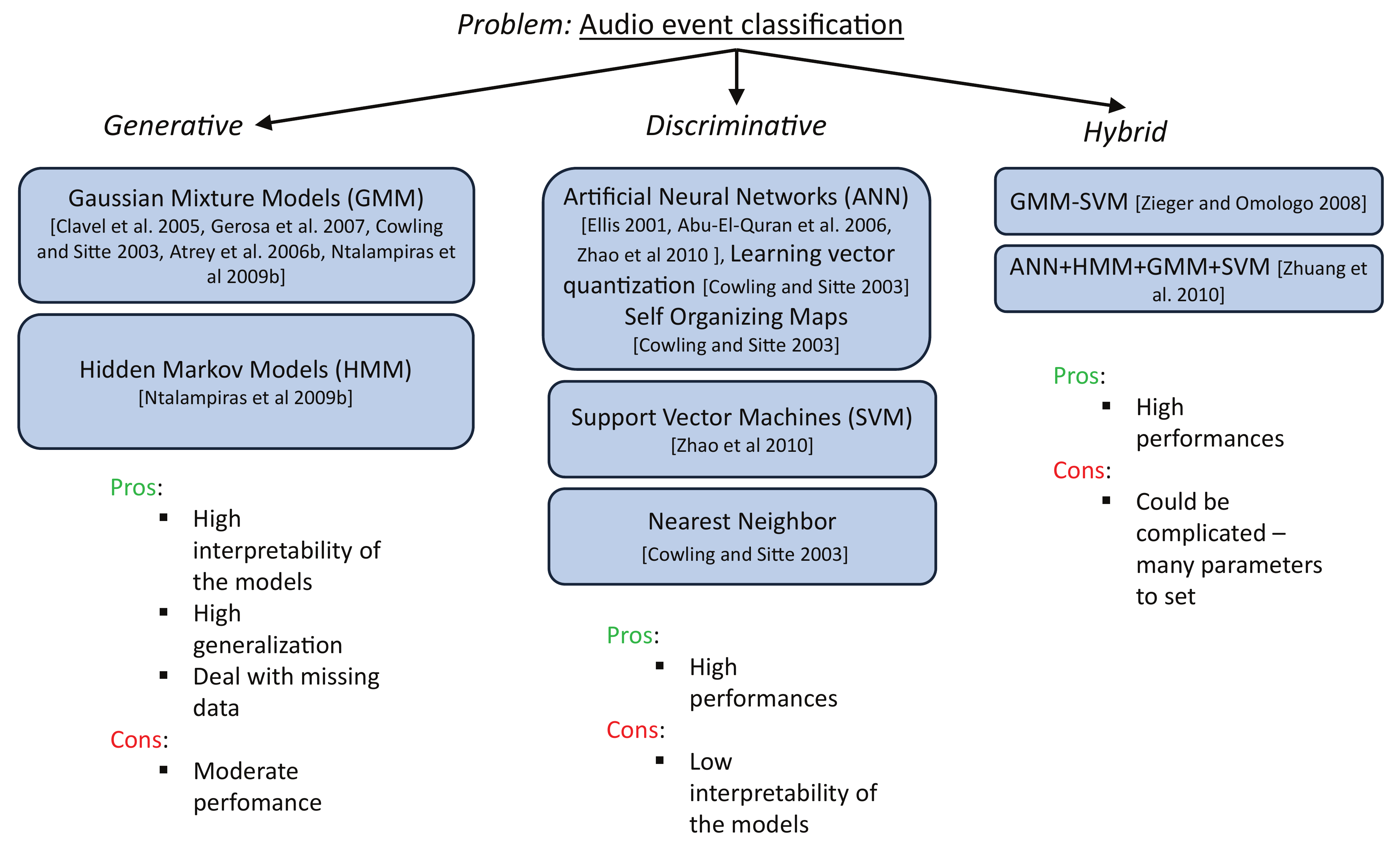}
\caption{Taxonomy for the classification methods, with pros and cons added for each category of approach.}
\label{fig:classification}
\end{figure}

The recognition of audio events depends usually on a classification strategy: first, features are extracted from class-labeled audio signals to learn a specific classifier in an off-line fashion; second, the trained classifier is employed to recognize unseen audio samples.
A simple taxonomy, displayed in Fig.~\ref{fig:classification}, subdivides classification methods in generative and discriminative;
in the former case, each class of audio events has its own classifier, trained on samples of the same class. Usually, generative classifiers are defined into a Bayesian context, so that the classification score assigned to a test sample is a posterior probability. Given a test sample, multiple classifiers are evaluated (one for each class), and the highest a posteriori probability determines the chosen classifier, and thus the chosen class. Among generative models Gaussian Mixture Models (GMM) and Hidden Markov Models (HMM) are the most widespread in the audio classification field. In particular, HMMs are suited to model the temporal variation of the feature vector over consecutive frames, allowing a more accurate modeling of each sound class. Transient sounds, such as gunshots or screams have typical temporal signatures which can be captured with left-right HMMs \citep{Rabaoui2009}, whereas stationary sounds can be efficiently modeled by ergodic HMMs.


On the other hand, discriminative models try to directly construct the best hyper-surface of separation in the feature space, dividing it into subspaces segregating the most training samples for each class. Artificial Neural Networks (ANN) and Support Vector Machines (SVM) are the most common discriminative models employed in the audio classification task.

Concerning audio event classification, the state of the art is far from conclusive toward a common framework as, for example,
for speech/speaker recognition where the classifier and the feature extraction process is rather established (i.e., GMMs and HMMs as classifiers and variations of spectral features as input \citep{Ntalampiras2009_1}).
In fact, the direct application of state-of-art techniques for speaker or musical instrument recognition to environmental sounds provides in general poor performance \citep{Cowling2003}. The challenge lies here on the fact that it is difficult to foresee all the kinds of sounds which could be present in a given environment, and often very little samples of unusual sounds are available to properly train a classifier.

Furthermore, differently from speech, generic sound events may have very different characteristics including duration, spectral content and volume with respect to background noise. Finally, the microphone(s) can be located far from the acoustic source, therefore implying strong echoes and reverberations (especially in indoor environments) and low SNR at the received signal. For these reasons, the findings from works not explicitly devoted to surveillance tasks \citep{Guo2003,Lin2005,Lu2002,Li2000} cannot be straightforward extended to the audio surveillance field, despite useful ideas can certainly be borrowed.\\

In audio surveillance, several works recently presented dealt with a limited set of sound classes. In \citep{Ellis2001}, an ANN is used to detect the presence of several alarm sounds, treated as a single class, against generic ambient noise. In such a case, audio classification closely resembles background subtraction, where foreground is predefined offline and encompasses a specific sound type.

A similar task was faced by \citep{Clavel2005}, in which the audio events of interest are gun shots. Two approaches are adopted: in the first one, a couple of GMMs is trained offline so as to model shot class and ``normal'' audio class; in the second one, several GMMs, one for each kind of shot (e.g. gun shot, rifle shot etc.), are trained offline and used to implement a series of binary classifiers (normal sound vs. specific shot) and the final decision on shot is taken if at least one of the classifiers detects a kind of shot. The latter approach allows to significantly improve the recall at the expense of a slight decrease in the precision. In \citep{Gerosa2007},
two classifiers based on GMMs run in parallel in order to detect respectively scream and gun shot against normal sound and the decision that an harmful event (either scream or gunshot) has occurred is taken computing the logical OR of the classifiers. 

A comparative analysis of several classifiers, including Learning Vector Quantization, GMM, ANN, Dynamic Time Warping and Long-Term Statistics, coupled with different features, was performed in \citep{Cowling2003}. The best results with a 70\% of samples correctly classified were obtained with Dynamic Time Warping, but the small size of training and test sets does not allow to draw a general conclusion.\\

If the number of classes increases, a hierarchical classification scheme, composed of different levels of binary classifiers, generally achieves higher performance than a single level multiclass classifier. Following this principle, in \citep{Atrey2006} five sound classes (talk, cry, knock, walk, run) are discriminated by a GMM classification tree whose intermediate nodes comprehend vocal events, non vocal events and footsteps. A similar approach was used in  \citep{Ntalampiras2009}.\\

Moving to the discriminative classifers, hierarchical approaches were followed in \citep{Zhao2010} using binary SVMs, and in \citep{Abu-El-Quran2006} using ANN. The same hierarchical scheme is employed in \citep{Rouas2006}, but in this case only two final classes shout/non-shout are considered. The tree is aimed at progressively excluding background noise, non-voice sound and non-shout voice, so yielding a consistent improvement in precision (lower false alarm rate) with respect to a single level classification. In \citep{Choi2012}, the first-level classifier subdivides sounds into harmonic ones and non-harmonic ones. The former are subsequently classified into voice and non-voice, and the latter are classified into low brightness (like glass breaking) and high brightness (like gunshots). \\
As SVM classification was originally developed for binary discrimination, the extension to multi-class classification has been achieved by a set of one-against-one or one against-all strategies: in the former case $N\cdot(N-1)/2$ SVMs, $N$ being the number of classes, are trained with data related to each couple of classes, while in the latter  case $N$ SVMs are trained taking into account all the data available. In both cases, the final classification is achieved by a voting procedure. An interesting alternative is reported in \citep{Rabaoui2008}, where $N$ One-Class SVMs are trained with data belonging to just one class for each SVM. In the testing phase, a dissimilarity measure is calculated between the current data sample and each 1-class SVM, and the sample is assigned to the class yielding the lowest dissimilarity value. Other than the computational advantage in the training phase, this approach provides a natural way to classify as unknown a given data. If the dissimilarity measure is higher than a predefined threshold for all One-Class SVMs, the data is classified as not belonging to anyone of the predefined classes.\\

The complementary strengths of generative and discriminative models can be exploited by hybrid strategies. For example, in \citep{Zieger2008} a GMM-SVM couple is istantiated for each class. Given a data sample a combined score is produced by each GMM-SVM couple by a weighted sum of normalized scores related to GMM and SVM, with weights inversely proportional to the classification error rate. Finally, the classification is performed on the basis of the highest combined score.
A more elaborate scheme is proposed in \citep{Zhuang2010} for joint segmentation and classification of an audio stream. The first stage, composed of an ANN and an HMM in cascade, provides segment boundaries and Maximum a Posteriori (MAP) probabilities for each class. The probabilities are used to train a GMM model whose parameters are finally fed to an SVM that provides a refined estimation of the class labels for each segment.\\
The recorded sounds of interest are normally superimposed to environmental and electronic noise which determines a given SNR. If the classification is aimed at distinguishing between background noise and a given sound type, the SNR in training phase can affect the trade-off between precision and recall in the testing phase: high SNR yields a statistically better precision at the expense of recall and \textit{vice-versa} as shown in \citep{Clavel2005}. Moreover, the change of SNR between training and testing phase can negatively impact on the overall classification performance. To overcome this drawback in \citep{Dufaux2000} several noise levels were superimposed to the training sound samples to build an array of classifiers (GMM or HMM), each one targeted to a particular SNR. In the testing phase, after a coarse SNR estimation the classifiers with the nearest SNR level was applied. A similar approach to the problem was pursued in \citep{Rabaoui2009} and \citep{Choi2012}, where a single classifier was trained replicating the data with different SNRs (the so called \textit{multi-style} training).
On Table~\ref{table:EventsAC2}, a summarizing scheme which focuses on the several different audio events (so far described), and related classification methods is reported. Moreover on Table \ref{table:Feat_AC} features employed for audio event classification and related references are displayed.

\begin{table*}[!htbp]
\centering
{ 
{\scriptsize
\begin{tabular}{| p{5 cm} |  p{6 cm}|}
\hline
 \textbf{Event typology} & \textbf{Reference}  \\\hline
 Alarms, sirens, klaksons   & Left-right HMM \citep{Li2009}, discriminative GMM \citep{Kim2011}, discriminative GMM \citep{Kim2011}, ANN \citep{Ellis2001}\\\hline
 Ambient   & Discriminative GMM \citep{Kim2011}, GMM \citep{Vu2006}\\\hline
 Applauding   & Hybrid GMM/SVM \citep{Zieger2008}, left-right HMM \citep{Li2009}\\\hline
 Beating (fight)   & Bayesian networks + KNN \citep{Giannakopulos2010}.\\\hline
 Car braking   & Left-right HMM \citep{Li2009}\\\hline
 Car engine, bus engine    & Left-right HMM \citep{Li2009}\\\hline
 Chair moving   & Hybrid GMM/SVM \citep{Zieger2008}\\\hline
 Cheer   & Left-right HMM \citep{Li2009}\\\hline
 Coins dropping   & GMM, HMM,  \citep{Cowling2003},Learning vector quantization, ANN, SOM \citep{Cowling2003}\\\hline
 Coughing   & Hybrid GMM/SVM \citep{Zieger2008}\\\hline
 Crashing   & Left-right HMM \citep{Li2009}\\\hline
 Crying   & GMM classification tree \citep{Atrey2006} GMM \citep{Ntalampiras2009_1}, \citep{Pham2010}, Binary SVM \citep{Zhao2010}ANN \citep{Abu-El-Quran2006}, Discriminative GMM \citep{Kim2011}\\\hline
 Door opening/slammiing & Hybrid GMM/SVM \citep{Zieger2008}, discriminative GMM \citep{Kim2011}\\\hline
 Footsteps   & GMM, HMM,  \citep{Cowling2003}, \citep{Atrey2006}, \citep{Menegatti2004} GMM classification tree \citep{Atrey2006}, \citep{Ntalampiras2009_1}, hybrid GMM/SVM \citep{Zieger2008}, discriminative GMM \citep{Kim2011},Learning vector quantization \citep{Cowling2003}, SVM \citep{Atrey2006},  ANN, SOM \citep{Cowling2003}, \citep{Abu-El-Quran2006} \\\hline
 Glass breaking   & GMM, HMM,  \citep{Cowling2003}, Learning vector quantization, ANN, SOM \citep{Cowling2003}\\\hline
 Gunshots   & GMM \citep{Clavel2005}, \citep{Gerosa2007}, Bayesian networks + KNN \citep{Giannakopulos2010}        \\\hline
 Jangling keys   & GMM, HMM,  \citep{Cowling2003}, hybrid GMM/SVM \citep{Zieger2008}, Learning vector quantization \citep{Cowling2003},  ANN, SOM \citep{Cowling2003}\\\hline
 Keyboard typing   & Hybrid GMM/SVM \citep{Zieger2008}\\  \hline
 Knock   & GMM classification tree \citep{Atrey2006} GMM \citep{Ntalampiras2009_1}, Atrey et al 2006, hybrid GMM/SVM \citep{Zieger2008}, Binary SVM \citep{Zhao2010}, ANN \citep{Abu-El-Quran2006}\\\hline
 Laughing   & Hybrid GMM/SVM \citep{Zieger2008}, left-right HMM \citep{Li2009}\\\hline
 Music    & Left-right HMM \citep{Li2009}, Bayesian networks + KNN \citep{Giannakopulos2010}\\\hline
 Paper rustling   & Hybrid GMM/SVM \citep{Zieger2008}\\    \hline
 Phone ringing   & Hybrid GMM/SVM \citep{Zieger2008}\\\hline
 Running   &  GMM \citep{Atrey2006}\\\hline
  Scream, Shouting   & GMM \citep{Gerosa2007}, \citep{Atrey2006} \citep{Vu2006}, hierarchical GMM \citep{Rouas2006}\\\hline
 Spoon/cup jingling   & Hybrid GMM/SVM \citep{Zieger2008}\\\hline
  Tag ticket   & GMM \citep{Vu2006}.\\\hline
 Talk,voice   & GMM classification tree \citep{Atrey2006} GMM \citep{Ntalampiras2009_1}, Bayesian networks + KNN \citep{Giannakopulos2010}, Binary SVM \citep{Choi2012}, one-class SVM \citep{Rabaoui2008}.\\\hline
 Walk   & GMM classification tree \citep{Atrey2006} GMM \citep{Ntalampiras2009_1}, \citep{Atrey2006}, left-right HMM \citep{Li2009},Binary SVM \citep{Zhao2010}, ANN \citep{Abu-El-Quran2006}.\\\hline
 Wood snapping   & GMM, HMM,  \citep{Cowling2003},Learning vector quantization, ANN, SOM \citep{Cowling2003}.\\   \hline
\end{tabular}
}
}
\caption{  Events typologies and related classification strategies adopted in the literature.}
\label{table:EventsAC2}
\end{table*}

\begin{table*}[htbp]
\centering
{ 
{\scriptsize
\begin{tabular}{| p{1.5 cm} | p{5 cm} | p{6 cm}|}
\hline
 \textbf{Class} & \textbf{Short Description} & \textbf{Reference}  \\
 Time   & Zero Crossing Rate (ZCR) &\citep{Valensize2007,Vacher2004,Rabaoui2009,Rabaoui2008,Istrate2006,Gerosa2007,Atrey2006,Choi2012}.\\
        &  Correlation-based features: Periodicity, correlation slope, decrease and roll-off & \citep{Valensize2007,Gerosa2007}.\\
        &  Pitch range features: calculated from short time autocorrelation function & \citep{Uzkent2012}.\\
        &  Waveform minimum and maximum & \citep{Ntalampiras2009}. \\
        \hline
 Frequency  &  Fourier coefficients & \citep{Zhao2010,Cowling2003}. \\
            &  Spectral moments & \citep{Valensize2007,Gerosa2007,Clavel2005,Choi2012}.\\
            &  Spectral slope and decrease & \citep{Valensize2007,Gerosa2007}.\\
            &  Spectral roll-off & \citep{Valensize2007,Vacher2004,Rabaoui2009,Rabaoui2008,Istrate2006,Gerosa2007,Choi2012}.\\
            &  Spectral flatness & \citep{Ntalampiras2009,Choi2012}.\\
            &  Spectral centroid & \citep{Vacher2004,Rabaoui2009,Rabaoui2008,Istrate2006}.\\
            &  Pitch Ratio & \citep{Abu-El-Quran2006}\\
            \hline
 Cepstrum   &  MFCC & \citep{Ramos2010,Lee2010,Kim2011,Zieger2008,Zhuang2010,Zhou2008,Zhao2010,Valensize2007,Vacher2004,Rouas2006,Rabaoui2009,Rabaoui2008,Istrate2006,Cowling2003,Clavel2005,Chu2009,Choi2012,Abu-El-Quran2006}.\\
            &  MFCC derivatives & \citep{Zieger2008,Kim2011,Zhou2008,Rouas2006,Istrate2006,Gerosa2007,Choi2012,Abu-El-Quran2006}.\\
            &  Homomorphic Cepstral Coefficients & \citep{Cowling2003}.\\
            &  Linear Prediction Cepstral Coefficients (LPCC) & \citep{Rabaoui2009,Rabaoui2008,Atrey2006}.\\
            \hline
Time-frequency  & Wavelet coefficients & \citep{Rabaoui2009,Rabaoui2008,Cowling2003}.\\
                & Discrete Wavelet Transform Coefficients (DWTC) & \citep{Istrate2006,Vacher2004}.\\
                & Mel Frequency Discrete Wavelet Coefficients (MFDWC) & \citep{Rabaoui2008}.\\
                & Gabor Atoms & \citep{Chu2009}.\\
                & Short Time Fourier Transform  & \citep{Hoiem2005,Cowling2003}.\\
                & Trace transform applied to spectrogram & \citep{Gonzalez2007}.\\
                & Visual features applied to spectrogram & \citep{Souli2011}. \\
                & Local Autocorrelation of Complex Fourier Values (FLAC) & \citep{Ye2010}. \\
                \hline
 Energy         & Signal energy  & \citep{Rouas2006,Clavel2005}.\\
                & Log energy first and second derivatives & \citep{Zieger2008}\\
                \hline
 Biologically or perceptually driven  & Log frequency filterbank & \citep{Zhuang2010,Zhou2008}.\\
                & Narrow Band Autocorrelation Features  & \citep{Valero2012}.  \\
                & Gammatone Cepstral Coefficients (GTCC) & \citep{Valero2012a}. \\
                & Linear Prediction Coefficients and derivatives (LPC) & \citep{Rouas2006,Cowling2003,Atrey2006,Choi2012}.\\
                & Perceptual Linear Prediction Coefficients and derivatives (PLP) & \citep{Rouas2006,Rabaoui2009,Rabaoui2008,Cowling2003}.\\
                & Relative Spectral (RASTA) Perceptual Linear Prediction & \citep{Rabaoui2009,Rabaoui2008}.\\
                & Intonation and Teager Energy Operator (TEO) based features & \citep{Ntalampiras2009}.\\\hline
\end{tabular}
}
}

\caption{ Summary features for audio event classification: first column indicates the feature class according to the taxonomy defined in Section~\ref{sec:audiofeatures}, second column reports feature names and third column reports references of works where they are employed.}
\label{table:Feat_AC}
\end{table*}

\section{Source localization and tracking}

\subsection{Audio Source Localization}

\begin{figure}[ht]
\centering
\includegraphics[width = 14cm]{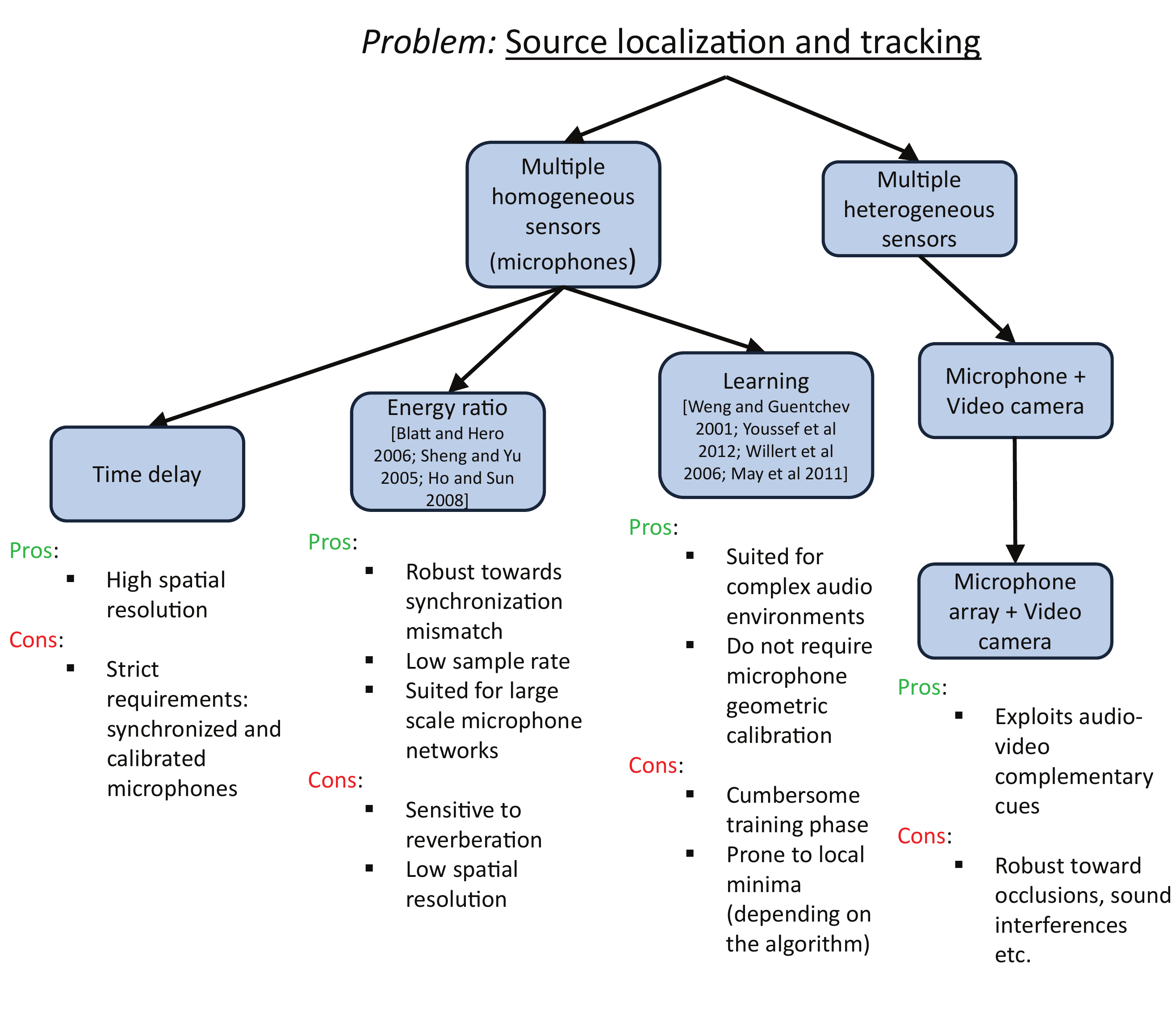}
\caption{General taxonomy of source localization: pros and cons are referred to energy-based (on the left) and learning-based methods (on the right).}
\label{fig:localization}
\end{figure}

When localizing a sound source, a single microphone samples the whole propagating wavefield in the scene, producing a one-dimensional electric signal as output. Therefore, differently from a video sensor, the spatial location of the source emitting the sound can not be inferred, unless knowing a priori the emitted signal, the environment characteristics, and the relation between the source location and the cues of the acquired signal, e.g. energy and spectral distribution. As an example, knowing the energy of the emitted sound and supposing to be in free space, one could estimate the source distance from the microphone by measuring the energy of the acquired signal. Moreover, if the microphone is not omnidirectional, also the sound direction of arrival could be in line of principle estimated. However, in a general case, the information on the spatial environment and the audio scene is quite limited or not available at all \footnote{Recently Acoustic Vector Sensors (AVS) have been imployed in a surveillance context \citep{Kotus2014}: differently from microphones a single AVS is able to measure the sound direction of arrival; however the diffusion of such kind of devices is to date quite limited.}. Therefore, it is mandatory to rely on multiple sensors, either homogeneous or heterogeneous. In the first case, a number of microphones is deployed in a given spatial configuration, obtaining a \textit{microphone array} or a \textit{microphone network}, in order to spatially sample the acoustic wavefield. From the set of acquired signals, a panoply of techniques \citep{VanTrees2002,Johnson1992} based on array signal processing can be applied to estimate the source location. In the latter case, the single microphone is associated to a natively spatial sensor, typically a video camera, trying to infer the sound source spatial location by exploiting temporal correlations between pixel changes in the visual image, likely caused by the object emitting sound, and sound cues. The two previous approaches can be combined together, fusing the spatial information provided by microphone arrays and video cameras to achieve an increased robustness and performance. A general taxonomy of source localization methods is summarized in Fig.~\ref{fig:localization}.\\

Concerning sound source location estimation by a microphone array, an essential taxonomy is reported here, describing the pros and cons of each methodology. A first partition individuates the  \emph{time-delay}, \emph{energy-ratio} and \emph{learning} based methods. The first class, by far the richest and most investigated, exploits the fact that signal delay of arrival is proportional, in free space, to the distance between each microphone and the sound source. The second one relies on signal energy attenuation which is inversely proportional to the square of the distance from the source. Finally, the last one does not assume a particular propagation model but tries to extract features from the set of audio signals in order to learn a regression function linking the feature vector and the sound location. \\
Time-delay-based methods can be in turn subdivided into three categories, as summarized in Fig.~\ref{fig:time_delay}: Steered Beamforming, High Resolution Spectral Estimation and Time Difference of Arrival (TDOA)  \citep{Brandstein1997}.\\
\begin{itemize}
\item \textbf{Steered Beamforming.} A beamformer or \textit{beamforming algorithm} denotes a technique devoted to spatially filter a wavefield originating from one or multiple sources. More specifically, the beamformer tries to attenuate as much as possible all the signals coming from different directions, while letting unaltered the signal coming from the direction of interest, known as \emph{pointing} or \emph{steering direction}. In its simplest version \citep{VanTrees2002}, known as delay-and-sum beamforming, the signal at each microphone is delayed in order to compensate for the propagation delay related to a given steering direction; after that, all the signals are summed together producing the beam signal, i.e. the beamformer output. The signal components related to the source in the steering direction will sum coherently, due to the delay compensation, while all the other components will sum uncoherently. Therefore, the beam signal will be representative of the signal of interest plus a residual sum of all the other signals attenuated.\\
Steered Beamformer based methods evaluate the beam signal energy on a grid of directions covering all the space of interest and search for the maxima which should correspond to the direction of arrival of the sounds. With linear microphone array, the more widespread one, localization is limited to a single angle, e.g. elevation, while with planar or volumetric arrays the direction of arrival can be estimated in terms of both azimuth and elevation. If sources are located in the near field of the array also their distance from the array can be estimated, allowing a complete 3D localization.\\
A set of variants to this method have been proposed, including  Filter-and-Sum beamforming \citep{Crocco2014}, where each signal is filtered by a predefined FIR filter, Phase Transform (PHAT) \citep{DiBiase2001}, consisting in a sort of frequency whitening aimed at improving robustness toward reverberation, and Maximum Likelihood reformulations \citep{Zhang2008}.
The main advantage of Steered Beamformer-based method is the robustness against environmental noise and reverberation, allowing acceptable performance even in complex scenes. The main drawbacks are the relatively poor spatial resolution and the significant computational cost, especially working with big arrays ($50-100$ microphones) and fine grid discretization.

\item \textbf{High Resolution Spectral Estimation.} This class of methods \citep{Choi2005,Chen2002} takes as input the cross-correlation matrix of the signals acquired by the array and directly extracts the directions of arrival or location of the signals, via autoregressive (AR), minimum variance (MV) or eigen-analysis techniques \citep{Johnson1992}. The main advantage of these methods is the high spatial resolution achieved in comparison with the Steered Beamforming techniques. However, a series of drawbacks, including the sensitivity to reverberation, the limited number of sources that can be localized at the same time and the need for long time windows over which the signals should be statistically stationary, make their use in the surveillance scenario quite limited.

\item  \textbf{Time Difference of Arrival Based.} In this class of methods \citep{Huang2001,Brandstein1997} the procedure is split in two steps. First, the Time Differences of Arrival (TDOA) at each couple of microphones are estimated, typically by peak search in cross-correlation functions; second, such TDOA are employed to infer the source positions, typically by geometric methods based on curve intersections or rank constraints of the matrix of microphone - sources distances. TDOA-based method are computationally not-demanding since the first step can be accomplished just on a subset of microphone couples, while the second one is intrinsically lightweight. Moreover just the estimated TDOAs have to be sent to a central processing unit, while the first step can be performed in a distributed manner close to each microphone couple. Finally, TDOA method can be adapted to work in complex environments, where a subset of microphones may be occluded with respect to the source, due to architectural barriers \citep{Crocco2012,Gaubitch2013}. Such features make TDOA-based methods particularly suited for microphone networks, i.e. a set of microphones deployed at considerable distance each other and without a predefined geometric layout. In a surveillance scenario microphone networks represent often a more practical and cost saving solution in respect to microphone arrays, where microphones are densely packed in a costly, ad-hoc built, single device with a specific geometric structure. 
 One issue related with microphone networks is the network geometric calibration: to this end, recent approaches have been developed, that allow joint source localization and network calibration \citep{Crocco2012}. The main drawback of TDOA methods is however the information loss intrinsic to the two-step procedure, which may result in a sub-optimal solution. Moreover, some of the previous methods are based on the minimization of nonlinear functions that are prone to local minima, especially in presence of multiple sources.

\begin{figure}[ht]
\centering
\includegraphics[width = 14cm]{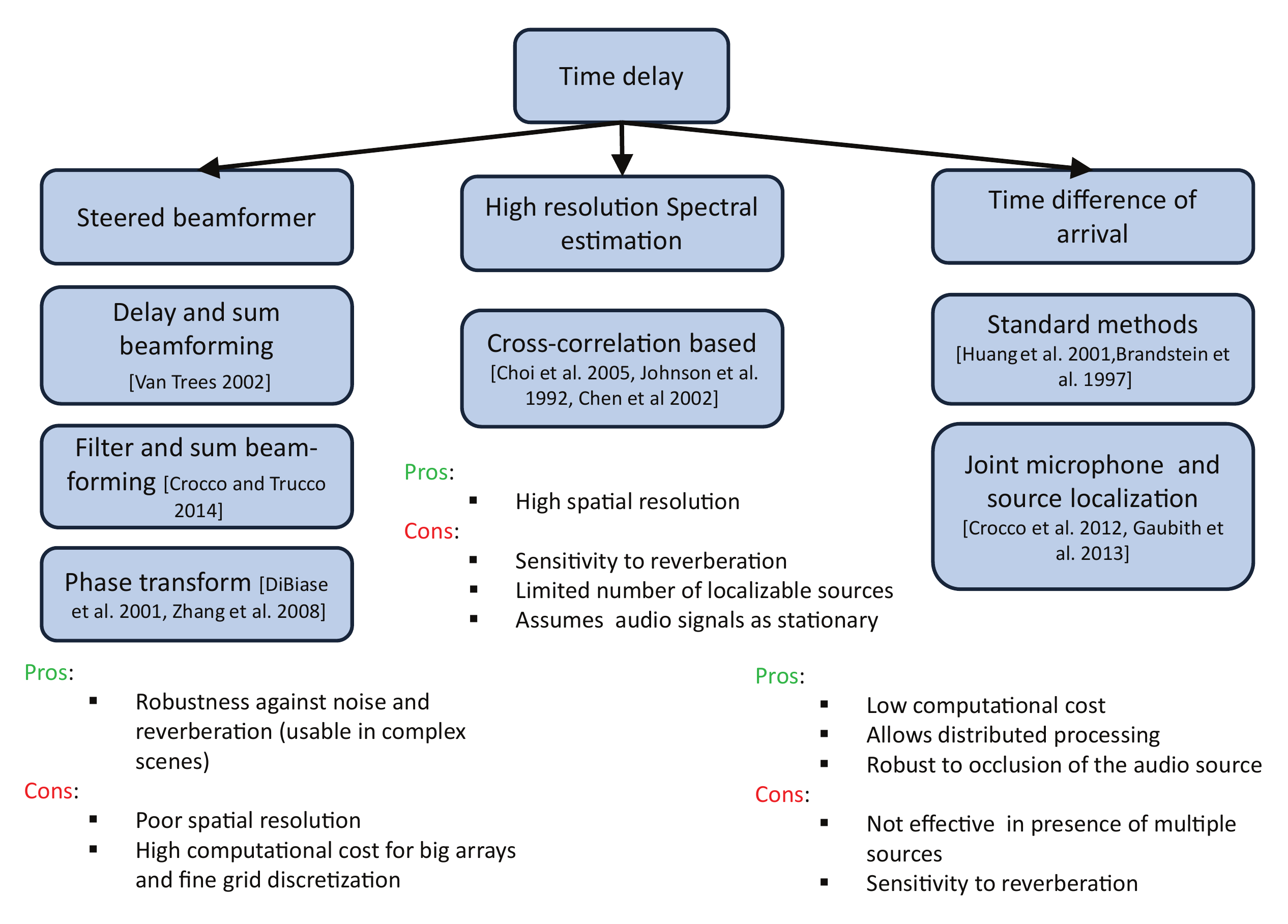}
\caption{Taxonomy for time-delay based localization methods}
\label{fig:time_delay}
\end{figure}

\item \textbf{Energy ratio based methods.} These methods \citep{Blatt2006,Sheng2005} are conceptually similar to the TDOA-based methods. In a first step, the signal energy ratio at each couple of microphones is evaluated; secondly the source location is estimated exploiting the relation between energy ratios and relative distances between microphones and source. Such techniques are generally adopted with microphone networks where the inter-microphone distances are sufficiently broad to allow substantial differences in the signal energies. Though their precision is generally inferior as compared to the three classes above based on propagation delays, and despite they are sensitivity to reverberation, some practical advantages make their use quite common in the surveillance context: energy evaluation does not require high sampling rates, so decreasing the burden of data transfer in wireless sensor networks, and is robust toward synchronization mismatch. Moreover, if sounds are narrowband, energy-ratio-based method can consistently improve the performance of TDOA-based methods \citep{Ho2008}.

\item \textbf{Learning Based methods.} This class of methods is based on a learning stage in which a set of features is extracted from the sound collected by the microphones and a classifier is trained in order to estimate the sound location. Such methods are comparatively less diffuse due to the difficulty in acquiring and annotating a reliable database of sounds encompassing all the range of possible locations, especially in uncontrolled environments. Moreover, the training phase needs to be repeated whenever the microphone array is moved in a different location. Nevertheless, learning-based methods allow to cope with complex environments implying strong reverberations, occlusions, nonlinear effects, deviations from the nominal parameters of the microphones, and in general all the deviations from the simple propagation and transduction model assumed by all the other methods. Moreover, local minima problems, arising in many of the previous methods, can be avoided by a learning strategy. The most common features adopted are Interaural Level Difference (ILD) and Interaural Time Difference (ITD), corresponding to energy ratio and TDOA among couple of microphones \citep{Weng2001}. A common approach \citep{Youssef2012,Willert2006,May2011}, inspired by the human auditory system, consists in preprocessing the signals with a gammatone filter bank, or cochleogram, mimicking the human cochlea frequency response, and subsequently extracting ITD, ILD and IPD (Interaural Phase Difference) on each single output of the filter bank. Such bio-inspired approaches typically work with just a couple of microphones and consequently limit the localization to the horizontal plane \citep{May2011}, unless elevation-dependent frequency distortion induced by the reflections of a synthetic human head is taken into account \citep{Youssef2012,Willert2006} \\
In Fig.~\ref{fig:2D_localization} a 2D diagram illustrates the optimal working conditions for each class of localization methods based on microphone arrays above described. Finally, in Table~\ref{table:Feat_SL} the set of features employed for audio localization is reported together with the related references.\\
\\
\end{itemize}

\begin{figure}[ht]
\centering
\includegraphics[width = 8cm]{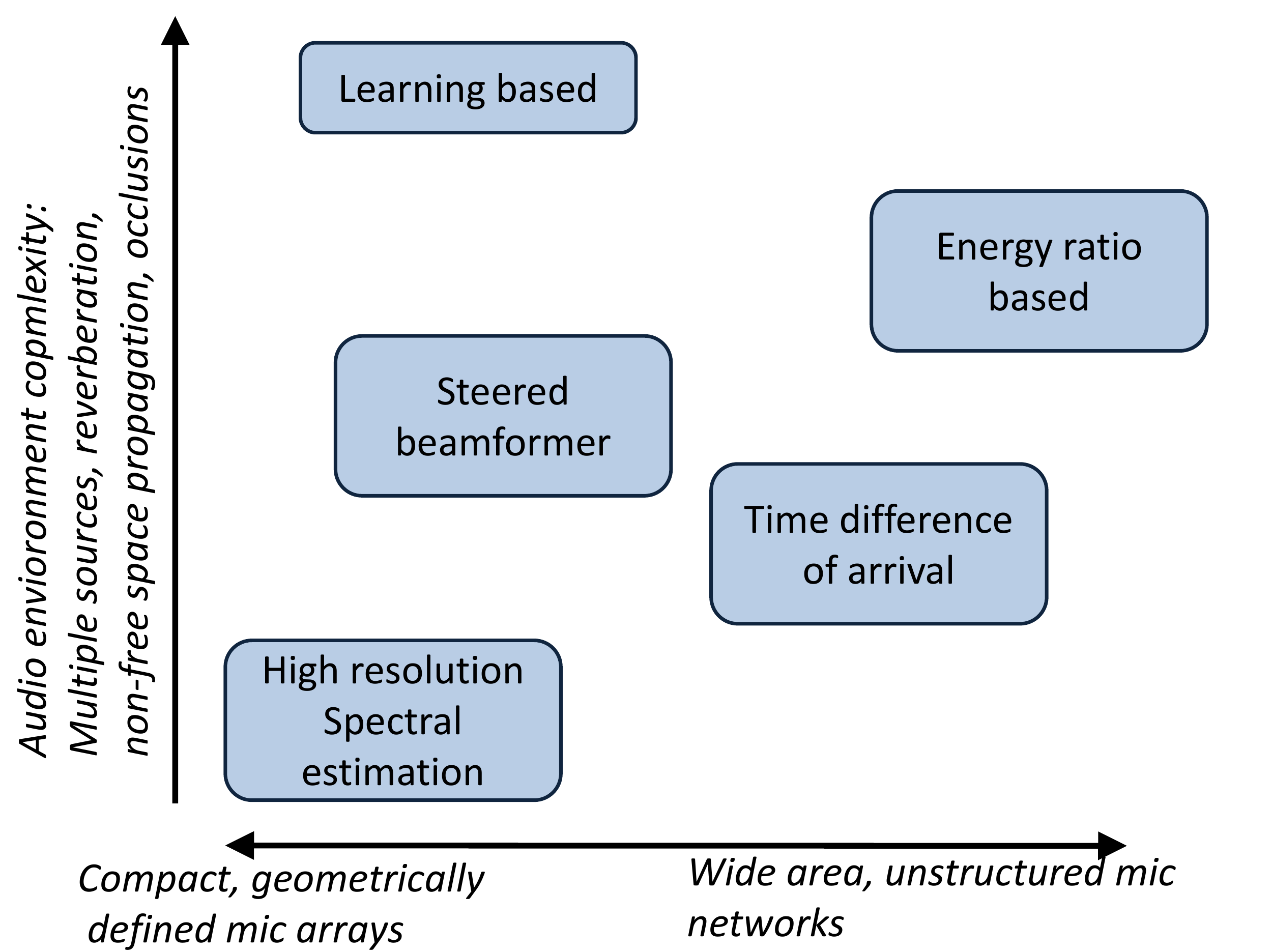}
\caption{Optimal working conditions for different localization methods.}
\label{fig:2D_localization}
\end{figure}

\subsection{Audio-visual source localization}

Single microphone localization techniques work in synergy with an optical imaging device: many environmental sounds (including voice, car engines, glass breaks, dog barks) have a visual counterpart, i.e. the image region depicting the sound source usually experience pixel changes that are temporally correlated with the sound emission (a clear example is given by a moving car). Exploiting such correlation it is possible to filter out the image background and the foreground not associated to the acquired sound, localizing in this way on the image the target sound source. A widely used methodology is Canonical Correlation Analysis (CCA) aimed at projecting audio and video signal onto a common subspace in which their correlation is maximized, while by inverting the process it is possible to recover the pixels associated to the sound source. CCA has been improved by imposing sparsity constraints \citep{Kidron2007} or working on intermediate feature representations \citep{Izadinia2013}  (MFCC for audio and pixel velocity and acceleration for video), rather than raw audio samples and image pixels.

In \citep{Hershey2000}, sound location is performed by looking at the image regions for which the mutual information between video stream and audio signal is maximized. In \citep{Smaragdis2003} Principal Component Analysis and Independent Component Analysis are applied in sequence to a compound vector of pixels and audio power spectra, so that the audio-video stream is segmented into independent components, each one corresponding to a different sound source. Another approach performs sparse-coding using joint audio-visual kernels \citep{Monaci2009} in order to learn bimodal informative structures from which the sound target location can be inferred. Finally, in \citep{Barzelay2010} a matching pursuit procedure is devised to localize multiple audio-video events: the adopted criterion is the temporal coincidence of audio and video onsets, the latter being evaluated by means of audio-visual features capturing strong temporal variations. The main limitation of these approaches is the audio-video matching ambiguity when multiple sounds and multiple moving objects occur at the same time. Moreover, they are unfeasible when no visual counterpart is present (e.g., a pipeline loss or a phone ring). It is interesting to note that the approaches above described can be considered as a particular kind of audio-video raw data fusion: the output of the fusion process is a pixel mask, function of time, selecting the ``sounding'' pixels.\\

\begin{table*}[!htbp]
\centering
  {
{	\scriptsize
\begin{tabular}{| p{1.5 cm} | p{5 cm} | p{6 cm}|}
\hline
 \textbf{Class} & \textbf{Short Description} & \textbf{Reference}  \\
\hline
 Time          & Interaural Time difference & \citep{Weng2001,Youssef2012,Willert2006,May2011}.\\
 \hline
 Frequency     & Interaural Phase difference & \citep{Youssef2012}.\\
 \hline
 Energy        & Interaural Level Difference & \citep{Weng2001,Youssef2012,Willert2006,May2011}.\\
\hline
\end{tabular}
}
}
\caption{ Features for audio localization: first column indicates the feature class according to the taxonomy defined in Section~\ref{sec:audiofeatures}, second column reports feature names and third column reports references of works where they are employed.}
\label{table:Feat_SL}
\vspace{-0.3cm}
\end{table*}

\subsection{Audio source tracking}

Tracking of an audio target can be performed in a naive way, simply updating the target position detected at each audio frame, according to the output of the localization algorithm. Such procedure is obviously not robust with respect to localization errors due to interfering sounds and is adopted only in the context of audio-visual tracking \citep{Beal2003}.\\

Differently, standard tracking algorithms take into account the dynamic of the source and recursively estimate the source location on the basis of the previous and current measurements or observations \citep{Arulampalam2002}. In particular, at frame $t$ the source location is predicted according to the evolution of the state dynamic, taking as initial condition the estimation at frame $t-1$ (prediction step). Subsequently, the predicted location is updated with the observation at frame $t$ (update step).

The most simple tracking algorithm is the Kalman Filter, which, under assumptions of Gaussian noise and linear functions wrt to the state for both dynamic and measurement, provides statistically optimal estimations. Unfortunately, such conditions are rarely fulfilled in the audio context. For example, in \citep{Strobel2001} the observation is given by the estimated azimuth angle and range provided by a microphone array, which are clearly nonlinear with the respect to the cartesian coordinates of the source location. For this reason, it is suggested to switch to the Extended Kalman Filter, based on local linearization of the observation function \citep{Strobel2001}.

In strong reverberation conditions, or in presence of impulsive disturbing sounds, the Gaussian assumption on the measurement noise does not hold anymore. In fact the observation function, given by a localization algorithm, will likely return a location quite far from the real source in a significant number of frames. In such a case, it is necessary to move to other algorithms able to handle arbitrary PDFs of the measurement noise. Among them, the Particle Filtering \citep{Arulampalam2002} has been widely used in the audio localization context \citep{Zotkin2002,Ward2003,Levy2011}. The underlying principle is to approximate the likelihood function, defined as the probability of obtaining the current observation from a given state, by a Monte Carlo simulation of a set of samples or particles. Each particle at time $t$ is weighted according to the observation, and the weights determine how many particles at time $t+1$ will be generated around each particle at time $t$. Finally, the estimated position is taken as the centroid of the particles. In this way particles are forced to crowd around the coordinates which are more likely to be close to the true target position. Moreover,  if a measurement is quite far from the previous estimation, probably it will fall in a particle-free zone and will not be taken into account in the update stage, so naturally filtering out reverberations and disturbing sounds.\\

\subsection{Audio-visual source tracking}

Despite the encouraging results demonstrated in ad-hoc and controlled setups, audio-only tracking methods have till now rarely been implemented in surveillance systems in real environments, due to the lack of robustness in complex sound scenarios. In such context, joining audio and visual devices allows each modality to compensate for the weaknesses of the other one, yielding a determinant boosting in the overall performance. For example, whereas a visual tracker may mistake the
background for the target or lose it due to occlusion, an audio-visual tracker could continue focusing on the target by following its sound pattern. Conversely, an audio-visual tracker could help where an audio-only tracker may lose the target as it interrupts emitting sound or is masked by background noise.

The main issue to be faced in audio-visual localization and tracking is the fusion of information of the single devices, which can take place at different processing levels: at the feature extraction level, simply \emph{feature level}, or at the \emph{decision level}. In the former case, a single tracker, typically a particle filter is instantiated and the fusion occurs when building the likelihood function. In particular in \citep{Aarabi2001} the final likelihood function is proportional to the sum of the two likelihood functions related to audio and video measurements, whereas in \citep{Zotkin2002} and \citep{D'Arca2013} the final likelihood function is proportional to the product. These two fusion strategies can be extended to multiple cameras and multiple microphone arrays \citep{Kushwaha2008}. More advanced fusion policies for the likelihood function are weighted sum \citep{Gerlach2012} or exponentiation of the two likelihoods with different coefficients prior to sum \citep{Gerlach2012}, in order to take into account the different reliability of the two modalities. In \citep{Beal2003}, fusion is implicitly achieved by a Bayesian Graphical Model in which observed variables, here two microphones signals and the video pixels, are modeled as depending from hidden variables denoting the target positions	and the audio and video noise (feature fusion level).

Considering the decision level fusion scheme, in \citep{Strobel2001}, audio and video data are processed separately by two independent Kalman filters; next the two localization outputs are fed to a fusion stage which provides a final joint estimation. Interestingly, the fusion stage can be conceptually divided into two inverse Kalman filters, recovering the audio and video measurements, and a joint Kalman filter that takes as input the audio-video measurement vectors and yields the final estimation.
 In \citep{Megherbi2005}, audio and video localization output are fused together using Belief Theory \citep{ayoun2001data}. Moreover, Belief Theory has the advantage to be able to handle the problem of associating multiple audio and video data to multiple targets.\\

Another issue raising in multimodal localization is the mutual geometric calibration of the different devices. In \citep{O'Donovan2007}, it is demonstrated that spherical microphone arrays can be considered as central projection cameras; therefore, calibration algorithms based on epipolar geometry, usually employed to calibrate a couple of video cameras, can be easily adapted to the problem of camera-microphone array calibration. In other works, the mutual calibration problem is jointly solved with the localization. In \citep{Beal2003}, calibration parameters and target locations are modeled together as hidden variables in a Bayesian framework, and inferred from the observations through Expectation-Maximization. In \citep{Zotkin2002}, calibration parameters are added to the state vector and tracked jointly with the target location, allowing to cope with unforeseen movements of the acquisition devices.\\

Beyond collaborative modality, audio and video devices can be exploited also in in a master-slave configuration. When the master role is played by the microphone array, the slave is usually a Pan Tilt and Zoom (PTZ) camera which is rotated and focused on the location from which an interesting or alarming sound has been emitted \citep{Bo-Wei2013,Kotus2013,Viet2013}. On the contrary, when the interest is focalized in hearing toward the direction where visual activity has been detected, the slave role is played by a microphone array \citep{Menegatti2004} which is electronically pointed, or simply a single directional microphone mechanically moved analogously to the PTZ camera.

\section{Situation analysis}
In respect to the previously described tasks, situation analysis, also known as scene analysis, deals with audio data at higher level of abstraction, trying to extract complex semantic concepts from the previous intermediate processing stages. Situation analysis generally involves the temporal and spatial integration of multiple data, often acquired from several heterogeneous sensors. As an example, a relevant situation for surveillance such as human aggression involves several agents, at least an aggressor and an attacked subject, holding anomalous behaviors (e.g., running, hitting, shouting), whose cues can be detected through audio and video modalities.\\

\begin{figure}[ht]
\centering
\includegraphics[width = 14 cm]{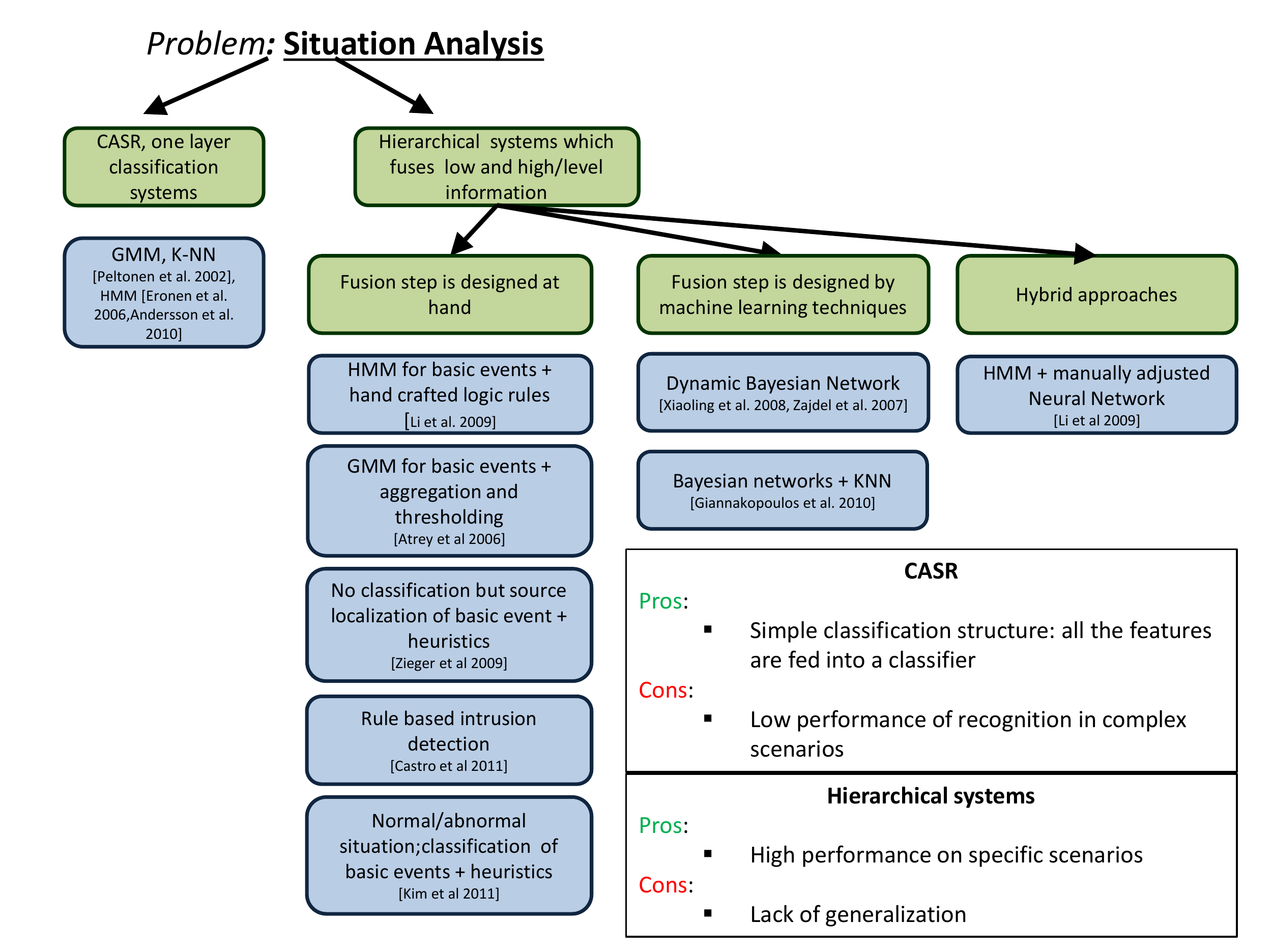}
\caption{Taxonomy for the situation analysis methods, with pros and cons added for each category of approach.}
\label{fig:sit_anal}
\end{figure}

\subsection{One-layer systems}
Though conceptually different from sound classification or foreground extraction, situation analysis can be faced using similar statistical approaches. The most straightforward one consists in defining a finite set of states characterizing a given environment, e.g. normal traffic, queue or accident on a road, or normal activity and aggression in a public space, and infer such states directly from a set of features extracted from the audio stream or audio-video stream through machine learning algorithms. According to this strategy, situation analysis is formally treated as a sound classification problem, where single sounds signals are substituted with more general scene descriptions. The approach was firstly addressed in the context of \emph{Computational Scene Recognition }(CASR) \citep{Peltonen2002} and \citep{Eronen2006}. Differently from the typical CASR application, i.e. a moving device recognizing the different environments crossed, in situation analysis each state is defined by a different situation related to a single environment where fixed sensors are deployed (e.g. normal traffic or car crash in the same route). The complexity of audio signatures related to each situation or environment is by far increased with respect to single sound sources, as very different sound may characterize alternatively or jointly a single situation (e.g., siren and crashing in a car accident), and the same sound can be shared by different scenes (e.g., sound of running people can occur either in a normal situation or in a threatening one). For these reasons performance is generally worse in comparison to sound classification task.\\

\subsection{Hierarchical systems}

A different approach exploits the inherent hierarchical structure of a scene, detecting at first time the single elements composing the scene, notably the single sound sources or the individual objects in the video case, and subsequently trying to fuse them in a second time according to a given policy. The fusion stage can be addressed either by 1) explicit rules incorporating human knowledge about the relationship among single sound events in a given scene, or 2) a machine learning approach.

An example of the first strategy can be found in \citep{Li2009}: first, basic audio events are modeled by a set of HMMs; second, audio events that are likely to occur simultaneously in an audio frame are grouped together defining \textit{a priori} the transition probabilities among them, and assuming that each basic audio event can belong to multiple groups, each one identifying a different structured audio scene.

In \citep{Atrey2006_2}, the detection of compound events is addressed fusing information coming from multiple sensors. In particular, three fusion levels are defined: \emph{media stream} level, \emph{atomic event} level and \emph{compound event} level. In the first one, features extracted from the data stream related to each sensor are stacked together (feature fusion level) and fed into a classifier in order to detect atomic events (e.g., walking, shouting, etc.). Second, each detector, related to just one sensor, yields a probability of detection for each atomic event. Such probabilities are fused together (decision fusion level) with a policy which takes into account both the different confidence of each sensor and the average level agreement of different sensors with respect to each atomic event. Third, probabilities of each compound event (e.g., a person running, while another one shouts) are estimated by fusing the probabilities of the subset of atomic events which define a priori the given compound event (decision level fusion). Detection of compound events is performed by a thresholding operation on the final probabilities. Such late thresholding policy allows to achieve higher accuracy, exploiting all the information available, in comparison to early thresholding which discards atomic events with low probability or sensors with low confidence.\\

An example of the second strategy, which does not incorporate prior knowledge, can be found in  \citep{Xiaoling2008}, where situation recognition is performed by a hierarchical Dynamic Bayesian Network (DBN), whose three hidden levels correspond, in descending order, to \emph{Situation event}, \emph{Group event} and \emph{Single events}, and the visible layer corresponds to audio and video cues. Since events at a given level can be regarded as the cues of the events at higher levels, all the statistical dependencies among different levels can be learned by a training procedure.\\

An hybrid approach, in which learned inferences are tuned according to prior knowledge, can be found in \citep{Li2009_2}. In this work, scene semantic content is extracted by means of a two-layer neural network. Single sound events are detected through HMMs, their duration and frequency in the audio sequence is used to calculate an input vector feeding the first layer of the neural network, corresponding to the single event level. Then the weighted sum of the inputs gives the probability of occurrence of each scene in the second layer. Finally, weights are adjusted on the base of the a priori judged importance of the related events for the given scene.\\

Differently from the vast majority of CASR applications where multiple audio contexts have to be recognized, scene analysis in the surveillance field is often devoted to distinguish among a normal situation and a specific relevant situation strongly related to a single environment. For example, detection of aggression in a public space \citep{Zajdel2007}, \citep{Andersson2010} or human intrusion in an private indoor space \citep{Menegatti2004}, \citep{Zieger2009} have been addressed in the literature.

More in detail, in \citep{Zajdel2007} an aggression detection system in a railway station has been devised based on a combined audio-video system. The audio part distinguishes normal speech from speech under strong stressing emotions, analyzing the pitch and the spectral tilt of the audio sequence. The video part tracks pedestrians and calculates their body articulation energy, which is used as a visual feature of aggression. A separate system detects passing trains in order to exclude false alarms. Audio and video cues of aggression are subsequently fed into a DBN which encodes the probabilistic dependency between the aggression level of the scene and the aggression cues (feature level fusion). The output of the DBN is the estimated time-dependent aggression level in the scene. \\

In \citep{Andersson2010}, a sensor set including two video cameras, a thermal camera and an array of microphones is employed to detect fights in outdoor urban environments. Audio and video stream are first processed separately: two HMMs modeling normal and abnormal audio events, together with audio features specifically targeted for abnormal vocalic reactions, are employed to reveal human sound associated to panic or aggression. Conversely, video streams, both visual and thermal, are used to estimate crowd size and activity. Information on crowd state and human voice are used as observations of a further HMM whose states model calm motions, or slightly increased activities (i.e., normal situations). Hence, low likelihood values for a given observation set denote abnormal, aggression-like situations.\\

In \citep{Giannakopulos2010}, audio and video features are extracted and fed separately into two Bayesian networks, whose outputs give the probability associated to two video classes (normal and high activity) and seven audio classes (including violent and non violent ones). Subsequently, such probabilities are considered as higher level features and fed into a nearest neighbor classifier, which yields to the final classification between violent and non violent activity. The fusion strategy in \citep{Andersson2010} and \citep{Giannakopulos2010} can be considered halfway between feature and decision level fusion, since the output of the first stage classification represents both the decision related to a given class (semantically different from the final ones) and the higher level features for the final classifier.

In \citep{Menegatti2004}, intruder detection in a dynamic indoor environment, e.g. the storage room of a shipping company, is addressed by means of audio and video static sensors and a mobile robot. The static cameras detect a moving object in the image communicating its position to the robot and the static microphone arrays. Microphones arrays are electronically steered toward the moving object and footsteps of the likely person are recorded and analyzed by a set of HMMs aimed at distinguishing between several known persons or an unknown one. The person is then tracked by microphone arrays and the information on the person location from both audio and video static sensors are fused together (feature fusion) and used to guide the robot toward the target person. 

In \citep{Zieger2009}, an heuristic strategy is devised to distinguish an actual intrusion in a room from false alarms generated by both ground noise coming from outside the room (road traffic, trains) or noise generated by static objects in the room (fridge pump, heating system). A network of microphones pairs is deployed in the room; each couple is able to measure the sound direction of arrival into a limited range. If a sound energy increase is detected,  but no definite direction of arrival can be measured by any of the microphone pair, the sound is discarded as it is likely to be diffused from outside the room. On the contrary, if a clear direction of arrival is measured, a counter is incremented. Each microphone couple can increment the counter just once, so that a sound produced by a static object in the pairs range causes just one increment of the counter. When the counter exceeds a predefined threshold, becoming higher than the maximum number of noisy objects in the room, an intrusion is detected.\\

Intrusion detection is also addressed in \citep{Castro2011} using an heterogeneous sensor network composed by microphones, video cameras and proximity sensors. The focus of this work is on the integration of mid-level single sensor output, such as people tracking, glass breaking detection, etc., by means of ontologies, fuzzy logic and expert systems, in order to get a semantic interpretation of the scene and trigger an alarm which also notifies its degree of confidence and other useful cues. \\

Other environment-specific methods concern security in public transportation, in particular small vehicles \citep{Kim2011} and trains \citep{Pham2010,Vu2006}. In \citep{Kim2011}, a system for detection of abnormal situations in small vehicular environments is proposed. A first processing stage classifies each audio frame into a given class drawn from two subsets of normal and abnormal events. Next, an abnormal situation is detected if the ratio of abnormal events in the whole audio sequence is higher than a predefined threshold.

The method proposed in \citep{Pham2010} is aimed not only at detecting an alarming event occurrence but also at identifying the person causing the event to happen. To this end, an audio-video sensor network is deployed in a train coach: a set of microphones located along the ceiling of the coach detects and locates alarming audio events such as shouts or spray bombs sending a warning to the human operator together with the image of the video camera closest to the audio event. After the human operator has selected the suspected person in the image, it is automatically tracked from video. When such person approaches a frontal camera, a further image is sent to the operator in order to allow face identification. A priori knowledge on the geometry of the environment is exploited for both audio localization and video tracking. As a matter of fact, the system involves the interaction with a human operator to identify the person to be tracked and cannot be considered truly automated.

\begin{table*}[!ht]
\centering
{  
{\scriptsize
\begin{tabular}{| p{5 cm} |  p{6 cm}|}
\hline
 \textbf{Event typology} & \textbf{Reference}  \\\hline
 Celebration & HMM combination \citep{Li2009}, \citep{Li2009_2}, Two-layer neural network \citep{Li2009_2}\\\hline
 Excitement & HMM combination \citep{Li2009}, \citep{Li2009_2} Dynamic Bayesian Network \citep{Zajdel2007}, Two-layer neural network \citep{Li2009_2}\\\hline
 Greetings & Dynamic Bayesian Network \citep{Zajdel2007}\\\hline
 HOME: bathroom & GMM \citep{Peltonen2002}, HMM \citep{Eronen2006}, K-NN \citep{Peltonen2002}\\\hline
 Normal VS abnormal & Discriminative GMM \citep{Kumar2005} \\\hline
 OFFICES/MEETING ROOMS/QUIET PLACES: Office, lecture, meeting, library, multi-person motion activity, multi-person speaking activity, using the projector, human presence activity, intrusion       & GMM \citep{Peltonen2002}, HMM\citep{Eronen2006}, \citep{Menegatti2004}, dynamic Bayesian Network \citep{Xiaoling2008}, K-NN \citep{Peltonen2002}\\\hline
 OUTDOORS: street, road, nature, construction site & GMM \citep{Peltonen2002}, HMM \citep{Eronen2006}, HMM combination \citep{Li2009}, \citep{Li2009_2}, K-NN \citep{Peltonen2002}, Two-layer neural network \citep{Li2009_2}\\\hline
 PUBLIC/SOCIAL PLACES: restaurant, caf\'e, supermarket & GMM \citep{Peltonen2002}, HMM \citep{Eronen2006}, K-NN \citep{Peltonen2002}\\\hline
 REVERBERANT: church, railway station, subway station & GMM \citep{Peltonen2002}, HMM \citep{Eronen2006}, K-NN \citep{Peltonen2002}\\\hline
 VEHICLES: car, bus, train, subway train, traffic  accident & GMM \citep{Peltonen2002}, HMM \citep{Eronen2006} HMM combination \citep{Li2009}, \citep{Li2009_2}, K-NN \citep{Peltonen2002},  Two-layer neural network \citep{Li2009_2} \\\hline
 VIOLENCE: Aggression, fight & Dynamic Bayesian Network \citep{Zajdel2007}, HMM Andersson et al 2010, Bayesian Network \citep{Giannakopulos2010}, K-NN \citep{Giannakopulos2010}\\\hline
 Vocal events VS non vocal event & GMM for basic events + aggregation and thresholding \citep{Atrey2006} \\\hline
\end{tabular}
}
}
\caption{ Situation typologies and related classification strategies adopted in the literature. Names in capital letters indicate scenarios where multiple situations have been taken into account with the same framework.}
 \label{table:SituationClass}
\end{table*}

\begin{table*}[!ht]
\centering
  {
{	\scriptsize
\begin{tabular}{| p{1.5 cm} | p{5 cm} | p{6 cm}|}
\hline
 \textbf{Class} & \textbf{Short Description} & \textbf{Reference}\\
\hline

Time      & Zero Crossing Rate  & \citep{Li2009_2,Eronen2006,Giannakopulos2010}.\\
          & Waveform minimum and maximum & \citep{Andersson2010}.\\
          \hline
Frequency &  Spectral centroid  & \citep{Li2009_2,Eronen2006,Clavel2008}.\\
           &  Spectral Roll-off & \citep{Eronen2006,Giannakopulos2010}.\\
           &  Band energy ratio(BER) & \citep{Li2009_2,Eronen2006}.\\
           &  Bandwidth & \citep{Li2009_2,Eronen2006}.\\
           &  Spectral Flatness  & \citep{Andersson2010}.\\
           &  Pitch Ratio & \citep{Giannakopulos2010}.\\
           &  Fundamental frequency & \citep{Andersson2010}.\\
           &  Spectral Variation  & \citep{Zieger2009}.\\
           &  Spectral Flux & \citep{Eronen2006}.\\
           \hline
 Cepstrum  & MFCC & \citep{Vu2006,Li2009_2,Li2009,Kumar2005,Eronen2006,Andersson2010,Clavel2008,Giannakopulos2010,Kim2011,Pham2010}.\\
           & MFCC derivatives & \citep{Vu2006,Eronen2006,Kim2011}.\\
           & Linear Prediction & Cepstral Coefficients and derivatives (LPCC) \citep{Eronen2006}.\\
           \hline
 Energy    & Signal Energy & \citep{Li2009_2,Zieger2009}.\\
           & Energy Enthropy & \citep{Giannakopulos2010}.\\
           \hline
 Biologically-Perceptually driven &  Pitch and spectral tilt  extracted from  filtered and thresholded cochleogram & \citep{Zajdel2007}.\\
                         &  Intonation and Teager Energy Operator (TEO) based features & \citep{Andersson2010}.\\
                         & Linear Prediction Coefficients and derivatives (LPC) & \citep{Eronen2006}. \\
                         & Prosodic group and Voice quality group & \citep{Clavel2008}.\\
\hline
\end{tabular}
}
}
\caption{ Summary of features for situation analysis: first column indicates the feature class according to the taxonomy defined in Section FIXME, second column reports feature names and third column reports references of works where they are employed.}
\label{table:Feat_SA}
\vspace{-0.3cm}
\end{table*}
On the contrary, in \citep{Vu2006} a fully automatic surveillance system based on audio-video sensor networks is proposed. The audio module detects alarming audio events, while the video module identifies and tracks all people and moving objects in the scene. The situation analysis is performed heavily relying  on a priori knowledge of the environment, including 3D geometry, static objects positions, physical properties and functionalities. Furthermore, a set of composite events is explicitly defined including the physical objects involved, the sub-events constituting the compound event, the sub-events not allowed during the compound event and a set of logical, spatial and temporal constraints among all events. Compound event detection is achieved by searching all the possible combinations of objects and sub-events detected, and checking whether the given combinations satisfy the above defined constraints. However, to avoid computationally unfeasible combinatorial explosion of sub-events, compound events are limited to the combination of at most two sub-events, this fact representing the major limitation of this promising approach.\\

A very specific task is addressed in \citep{Kotus2013}, i.e. the detection of kidnapping from a vehicle of a given person (supposed to be a ``Very Important Person''). The framework is very preliminary and implies an extended network of heterogeneous sensors, including a bluetooth wearable device which triggers the alarm when removed from the car, thermal and video cameras monitoring the scene and acoustic devices which detect and localize shout and abrupt impulsive sounds like gunshots. An alarm is finally delivered to a human operator joining all the multimodal cues (decision level).\\

Another ambitious work, presented as a proof of concept, is described in \citep{Kumar2005}. Here microphones,  video cameras and thermal cameras are jointly employed to infer several threatening situations. Video and thermal cameras perform segmentation and tracking of moving person/object; audio stream from a microphone is classified into a set of predefined classes, and finally semantic rules are applied to discover threatening situations like intrusion, theft, abandoned object (decision level). However, quantitative results are displayed only for single system modules.\\

An overall summary of the taxonomy for situation analysis methods is displayed in Fig. \ref{fig:sit_anal}. On Table~\ref{table:SituationClass}, a summarizing scheme which focuses on the several different audio situations (reported in the papers so far described), and related classification methods is reported. Finally in Table \ref{table:Feat_SA} the set of features employed for situation analysis is reported together with the related references.\\
\section{Audio Features}\label{sec:audiofeatures}
As mentioned in the introduction, a plethora of audio features has been devised in the last decades in the field of sound detection and classification. A large part of them was developed for specific tasks such as speech recognition \citep{Hunt1980}, speaker recognition \citep{Reynolds1994}, music classification \citep{Tzanetakis2002} or music/speech discrimination  \citep{Scheirer1997}, and a relevant subset has been later transferred in the audio surveillance context.  To guide the reader in such vast feature landscape, we rely on the comprehensive survey recently proposed in \citep{Mitrovic2010}. In this survey, a taxonomy is devised to describe audio features in a task-independent manner. Here, we focus on the feature subset actually employed in audio surveillance, detailing their use in the four macro-tasks so far discusses (background subtraction, audio events classification, source localization and tracking, situation analysis), and discussing several related issues such as computational load, expressivity, redundancy, robustness to noise, and others.\\

Generally speaking, audio features are aimed at encoding a high dimensional signal (typically an audio frame is made of hundreds or even thousands samples) into a small set of values that encapsulate the information useful for the detection or classification task, while discarding noise and redundancies. The taxonomy adopted in this survey subdivides audio features into six classes, namely \emph{temporal, spectral, time-frequency-based, cepstrum-based, energy based} and \emph{biologically or perceptually driven}.

The classification criterion for the first four classes is based on the intermediate signal representation from which features are extracted. In particular, temporal features are directly extracted from the signal samples, or more generally from a time domain representation of the signal, like the autocorrelation function. Spectral features are extracted from the signal spectrum, typically from its power modulus. Time-frequency features are extracted from a bidimensional representation function of time and frequency, like spectrogram, or time and scale, like wavelets. The 2D representation provides a rich amount of structure, useful for example to look at the spectral variation along time. Moreover, the 2D representation of the signal has recently encouraged the application of features borrowed from the image analysis domain. Cepstrum-based features are grounded on cepstrum, a nonlinear transformation of the spectrum which allows to compactly represent the spectrum envelope, discarding fine variations across close frequency bins (for example, the exact location of the harmonics in a periodic signal). Energy-based features deserve a separate class since their calculation is typically not associated to a given signal representation (energy can be extracted from time signal, spectrum, cepstrum and so on). Energy-based features are more involved in foreground extraction or tracking task, whereas they tend to be discarded for classification since generally they cause an increase in intra-class variation. Finally, biologically or perceptually driven features represent a class orthogonal to the previous ones: they can be grounded on temporal, spectral, time-frequency or cepstral representations but share a common inspiration from the psychophysiology of the human auditory and/or vocal apparatus. In particular, they can be subdivided in three subclasses: 1) features mimicking the processing of the human auditory system, in particular the cochlear filtering; 2) features reproducing the psychological perception of auditory cues and 3) features built according to the physical behavior of the human vocal tract, the latter being limited to encode vocal sounds.
Since the same audio features are often shared by the four macro tasks defined in the previous sections, we preferred to unify their description in the following. However, a set of four tables is displayed for each of the four tasks in the corresponding paper section. In each table, just the features employed for the specific task are reported, organizing them by the taxonomy above defined and reporting for each item the related references.\\
 Finally, a large amount of the features here described have been previously gathered into two large corpora of audio features aimed at encoding generic sounds: the two corpora are related MPEG-7 standard \citep{Chang2001} and CUIDADO project \citep{Vinet2002} respectively. In this review the possible inclusion of each feature in these corpora will be highlighted.

\subsection{Time}
\begin{itemize}
\item \textbf{Zero Crossing Rate (ZCR)}:  number of times the sign of the signal changes in a given frame. It captures information about the dominant frequency within the frame. Strongly correlated with the spectral centroid. Used in combination with more complex features (e.g. MFCC, Wavelets). Size = 1.
\item  \textbf{Autocorrelation coefficients} : temporal samples of the autocorrelation function calculated convolving the signal by itself. Autocorrelation is often employed as an intermediate representation from which other features can be extracted (see below). Size depends on the autocorrelation window.
\item \textbf{Correlation-based features}: Periodicity, correlation slope, decrease and roll-off.  Periodicity is calculated as the maximum local peak of normalized signal autocorrelation. Useful to discriminate between periodic signals (e.g. periodicity of a sinus wave = 1) and aperiodic signals (e.g. periodicity of white noise = 0). Slope, decrease and roll-off are similar to the corresponding spectral features but calculated from the autocorrelation rather than the spectrum. They are
suited for describing the energy distribution over different time lags. For impulsive noises, like gunshots, much of the energy is concentrated in the first time
lags, while for harmonic sounds, like screams, the energy is spread over a wider range of time lags. Size of each feature = 1.
\item  \textbf{Pitch range Features}: features extracted from the autocorrelation function evaluated on subsequent time windows. From each autocorrelation function, the pitch is estimated as the inverse of  the delay between the first and second highest positive peaks. Pitch range is then calculated as the ratio between the maximum and minimum pitch or the ratio between pitch standard deviation and pitch mean value. Pitch information is complementary to MFCC that encodes spectrum envelope; therefore their joint use seems to boost considerably the performance, especially on non-speech sound classification \citep{Uzkent2012}. Size = 2;
\item  \textbf{Waveform minimum and maximum}:  maximum and minimum value of the signal waveform. Included in MPEG-7 audio standards. Size = 2.
\item  \textbf{Interaural Time Difference (ITD)}: difference in the time of arrival of an acoustic signal at a couple of microphones. Used mainly for localization and tracking tasks. Size = 1.
\end{itemize}

\subsection{Frequency}
\begin{itemize}
\item \textbf{Fourier coefficients}: coefficients of the signal DFT typically averaged in squared modulus over subbands. Baseline feature in the spectral domain. Computationally efficient using Fast Fourier Transform. Size = number of subbands.
\item \textbf{Band Energy Ratio (BER)}: energy of the frequency subbands normalized by the total signal energy. Size = number of subbands.
\item  \textbf{Spectral moments}: statistical moments of the power spectrum, this last considered as a probability density function: spectral centroid or mean value (strongly correlated with ZCR), spectral spread (variance), spectral skewness, spectral kurtosis. Size of each one = 1.
\item \textbf{Bandwidth} : signal bandwidth, typically estimated as the frequency range in which a certain percentage of energy lies. Similar to spectral roll-off. Size = 1.
\item \textbf{Spectral slope, spectral decrease and spectral tilt}: spectral slope is defined as the amount of decreasing of the spectral amplitude with frequency. Computed by linear regression of spectral amplitude. Spectral decrease is the perceptual version of spectral slope. Spectral tilt is similar to spectral slope but is calculated as the ratio of energies in the spectrum lower portion (e.g. above 500 Hz) and higher portion \citep{Zajdel2007}. Included in CUIDADO corpus. Size = 1.
\item \textbf{Spectral roll-off}: frequency below which a certain amount of the power spectrum lies (tipically 95\% percentile ).  Useful to discriminate between voiced and unvoiced speech.  Size = 1.
\item  \textbf{Spectral flatness}: measure of noisiness/sinusoidality of the signal. Calculated as the ratio of the geometric mean to the arithmetic mean of the spectrum energy. Size = 1.
\item  \textbf{Pitch Ratio or Harmonicity}: in a fully periodic signal the spectrum energy is located at the fundamental frequency (the pitch) and its multiples (the harmonics), whereas in an aperiodic signal energy is spread over all the signal band. Such feature measures the degree of periodicity in a signal. Size = 1.
\item \textbf{Fundamental Frequency}: in a periodic signal corresponds to the inverse of the signal period. For a general signal is correlated with spectral centroid and ZCR.  Included in MPEG-7 corpus. Size = 1.
\item \textbf{Interaural Phase Difference (IPD)}: difference in the phase of the spectra of two signals acquired by a couple of microphones. If the two signals are generated from a sigle audio source, phase difference is proportional to the delay in the sound time of arrival at the two microphones. Mainly uesd for localization and tracking tasks. Size = number of frequency bins.
\end{itemize}
Most of the previus spectral features, modeling spectrum shape, are used in combination with higher size features (e.g. MFCC, DWT ). Poor performance if taken alone due to intrinsic low dimensionality. 

\subsection{Cepstrum}
 \begin{itemize}
 \item \textbf{Mel Frequency Cepstral Coefficients (MFCC)} :  the Cepstrum is the Discrete Cosine Transform of the log-magnitude of the signal spectrum. The Mel-cepstrum is calculated on the outputs of a Mel-frequency filter bank rather than directly on the spectrum. Mel-frequency filter bank is built according to the Mel scale, a nonlinear frequency scale corresponding to perceptually linear pitch scale, and characterized by a linear portion at low frequencies and a logarithmic portion at high frequencies.  MFCC are the first coefficients of the Mel-cepstrum (typically excluding the first one that accounts for the signal energy). MFCC provide a compact representation of the spectral envelope (formant structure), discarding fine cues like positions of the spectral peaks (harmonic structure).
Originally proposed in the context of speech-speaker recognition and successfully adapted to environmental sound recognition. Not suited for modeling noise-like flat spectrum sounds, like rain or insects. Higher order coefficients such as above $12$ are believed to contain more information about environmental sound sources other than speech \citep{Kim2011}. Often used in combination with simpler spectral and temporal features (ZCR, spectral centroid, roll-off etc.). Low robustness to narrowband noise due to DCT transform that spreads noise over all the coefficients. Moderate computational complexity. Size = number of selected coefficients; typically from $3$ to $15$.
\item  \textbf{MFCC derivatives}: derivatives of MFCC across adjacent temporal frames. Size = like MFCC.
\item \textbf{Homomorphic cepstral coefficients}: like MFCC but logarithm and discrete cosine transform are applied directly to the spectrum coefficients, without applying Mel filter-bank. Performance seems to be generally inferior to MFCC \citep{Cowling2003}. Size: like MFCC.
\item \textbf{Linear Prediction Cepstral Coefficients (LPCC)}: representation of Linear Prediction Coefficients (LPC) in the cepstral domain (see LPC). Size: like MFCC.
\end{itemize}

\subsection{Time-frequency}
\begin{itemize}
\item \textbf{Short Time Fourier Transform or Spectrogram}  \citep{Hoiem2005,Cowling2003}. Fourier transform applied to subsequent (possibly overlapped frames). PCA is used in  \citep{Cowling2003} to reduce feature space dimensionality. In \citep{Hoiem2005} STFT is used as intermediate representation in order to extract more synthetic features  (e.g. mean std of each frequency channel, bandwidth, most powerful frequency channel, number of peaks over time etc.).
\item \textbf{Wavelet coefficients}: Wavelet transform produces a joint time-frequency representation of the signal, allowing frequency dependent time resolution. This allows frequency
to be identified as occurring in a particular
area of the signal, aiding understanding of the
signal. Wavelet coefficients as well as the other time-frequency features have in general high cardinality. To reduce the feature space PCA is usually applied \citep{Cowling2003} \citep{Rabaoui2008} or coefficients are pooled over time (e.g. extracting mean, std, number of peaks) discarding in this way the temporal structure of the signal \citep{Rabaoui2008,Rabaoui2009}.
\item \textbf{Discrete Wavelet Transform Coefficients (DWTC)}: logarithm of the energies of last wavelets is evaluated and Inverse DWT is applied. Conceptually similar to MFCC but DFT and DCT are substituted with DWT and Inverse DWT. The final representation is not time-frequency. Wavelet is used as intermediate transformation. Better performance in respect to MFCC for low SNR conditions. Size = number of wavelet bases considered (typically six \citep{Istrate2006,Vacher2004}).
\item \textbf{Mel Frequency Discrete Wavelet Coefficients (MFDWC)}: like MFCC with the final DCT substituted with a Discrete Wavelet Transform. (wavelet transform applied to log magnitude of Mel filterbank). More robust to narrowband noise in respect to MFCC as only a subset of the final DWT coefficients is affected be noise.
\item \textbf{Mean and Standard Deviation of Gabor Atoms} Signal is approximated with a combination of Gabor functions via Matching Pursuit. Frequency and scale of each atom is evaluated and mean and standard deviation of frequency and scale are finally extracted. Size = 4 \citep{Rabaoui2009}\\
\item \textbf{Spectral Variation and Spectral Flux}: amount of variation in the spectrum across two adjacent time windows. Calculated as the generalized cosine (spectral variation) or the $L^2$ norm of the difference (spectral flux) between the two vectors representing the two spectra. Size = number of time windows couples taken into account.
\item \textbf{Trace transform applied to spectrogram}: spectrogram is treated as an image and Trace Transform is applied calculating integrals over lines parametrized by angle and distance from the image center. Final features are extracted calculating line integrals over diagonals of the transformed image. It seems to improve over MFCC \citep{Gonzalez2007} in general sound classification task. Trace transform is usually applied to affine invariant images therefore it could be useful to remove differences due to temporal or pitch translations.
\item \textbf{Visual features applied to spectrogram}: spectrogram is considered as a texture image on which visual features are extracted. In particular, scale and translation invariant wavelet transform is applied to the spectrogram; next, after a local max pooling a patch transform is applied and followed by a further max pooling \citep{Souli2011}.
\item  \textbf{High-order Local Auto-Correlation (HLAC)} : two dimensional autocorrelations calculated on the amplitude values of the spectrogram. Only local correlations ( $3 \times 3$ time-frequency lags) are taken into account. Feature borrowed from image classification field. Better results in respect to cepstrum based features in abnormal sound detection \citep{Sasou2011}.
\item \textbf{Local Autocorrelation of Complex Fourier Values (FLAC)}: similar to HLAC but complex values of the spectrogram, instead of amplitude only are taken into account.  Complex values allow to retain sound phase information. Definitely better performance in respect to MFCC in a classification task related to health monitoring \citep{Ye2010}.
\end{itemize}

\subsection{Energy}
\begin{itemize}
\item \textbf{Signal energy}: signal energy over a frame. Size = 1.
\item \textbf{Log energy first and second derivatives}. Size = 1..
\item  \textbf{Energy Entropy}:  Entropy calculated over the energy values extracted from a set of audio frames. Useful to detect abrupt energy changes, Size = 1 \citep{Giannakopulos2010}
\item \textbf{Interaural Level Difference (ILD) or Gain Ratio Of Arrival (GROA)}: ratio of the energies related to the signal acquired by a couple of microphones. Used mainly for localization and tracking tasks. Size = 1.
\end{itemize}

\subsection{Biologically/perceptually driven}
\begin{itemize}

\item \textbf{Log frequency coefficients}: outputs of a set of band-pass filters whose central frequencies and bandwidths are scaled according to a logarithmic scale. The logarithmic scale roughly reproduces the filter-bank response of the human ear. Typically used in combination with some feature selection process (e.g. Adaboost). Size = number of filters.
\item \textbf{Mel frequency coefficients}: similar to Log Frequency Coefficients, but the filters central frequencies are spaced according to Mel scale that reproduces the psycho-acoustical perception of the pitch. Mel scale is linear at the lowest frequencies and logarithmic at the highest ones. Mel scale is also adopted in several frequency, time-frequency or cepstral features. Size = number of filters
\item \textbf{Spectral features based on Gammatone filter bank}: this filter-bank aims at modeling the frequency analysis of the cochlea in the inner ear. Each filter has constant unitary bandwidth on an ERB frequency scale (Equivalent Rectangular Bandwidth). The energies of the filter outputs can be seen as a biologically driven version of the power spectrum from which spectral features are extracted. Size = number of filters.
\item \textbf{Gammatone Cepstral Coefficients (GTCC)}  variation of the MFCC in which the triangular filter bank spaced on the Mel scale is substituted with a Gammatone filter bank \citep{Valero2012a}, achieving better resolution at the lowest frequencies and tighter model of the human cochlea response. Performance comparable to MFCC (when used with SVM ) or significantly higher (when used with K-nearest neighbor) on non speech classification task. Size: like MFCC.
\item \textbf{Cochleogram}: biologically inspired version of the spectrogram. It is based on modeling of basilar membrane oscillations in human ear cochlea. The short time spectrum is calculated by a leaky integration of the squared membrane displacements \citep{Zajdel2007}. As for the spectrogram, it is normally employed as an intermediate time-frequency representation from which more synthetic features are extracted.
\item  \textbf{Linear Prediction Coefficients and derivatives (LPC)}: filter coefficients in an all-pole model approximating the audio signal. Originally devised to model the response of the human vocal tract (also known as Vocoder). Mainly targeted to speech vocal signals. Not suited for impulsive environmental sounds. Information captured is similar to MFCC. Size = typically some tens of coefficients.
\item \textbf{Perceptual Linear Prediction Coefficients and derivatives (PLP)}: mapping of LPC to the nonlinear frequency scale of the human ear. Information captured is similar to MFCC. Size = typically some tens of coefficients.
\item \textbf{Relative Spectral (RASTA) Perceptual Linear Prediction}: variation of PLP robust to background noise. The idea is to filter out stationary or slowly varying spectral components before the calculation of the all pole model (human auditory system pay little attention to steady state noise).  \citep{Hermansky1994}. Size = like PLP.
\item \textbf{Intonation and Teager Energy Operator (TEO) based features} : nonlinear operators, based on combination of signal derivatives,
that capture the variations that the vocal tract airflow exhibits when it comes to abnormal circumstances. Useful to distinguish threatening from normal situations on the base of human voice.
\item \textbf{Prosodic Group and Voice Quality Group}: a feature set tailored to human voice salient cues \citep{Clavel2008}. Prosodic Group includes Pitch, Intensity Contour and Duration of Voiced Trajectory; Voice Quality Group includes jitter (pitch modulation), shimmer (amplitude modulation), unvoiced rate (proportion of unvoiced frames in a sequence), and Harmonic to Noise Ratio (HNR) (proportion of periodic vs non-periodic signal).
\item \textbf{Narrow Band Autocorrelation Features}:  Features extracted from narrow band autocorrelation functions. Signal is firstly filtered by a filter bank whose central frequencies are spaced according to the Mel scale (mimicking the perceptual pitch). An autocorrelation function is calculated for each filter output and four features are extracted from each autocorrelation. The four features are the intensity of the first positive peak, the time delay between first and second peak, the intensity of the second positive peak and the duration of the envelope at $-10$ dB. They are related to four primary perceptual qualities of the sound, respectively loudness, pitch, timbre and duration.
Overcomes both MFCC and DWT features on a small scale classification task of non-speech signals \citep{Valero2012}. Size = 4 times size of filterbank.

\end{itemize}

\subsection{Feature selection and feature learning}

The current state-of-art in audio surveillance, and more in general in pattern recognition problems applied to audio signals, does not allow to draw an ultimate conclusion on the best feature or the best feature set to be used in detection and classification task irrespective of the kind of audio sources involved. The naive idea of stacking together as much features as possible, in order to boost the performance, clashes against the well known curse of dimensionality problem. To solve this issue the dimensionality of the feature space can be reduced with standard techniques like PCA or Independent Component Analysis \citep{Cowling2003}, or a feature selection process can be istantiated. To this end, in
\citep{Zhou2008,Zhuang2010,Hoiem2005} Adaboost meta-algorithm is employed to select an optimal subset of features. In detail, each single feature is associated to a weak-learner, and the final strong-learner output is given by a weighted sum of the weak-learners outputs. After the training phase, selection of the best $N$ features is performed simply looking at the highest $N$ weights. The rationale is that if a weak learner is associated to a high weight, the corresponding feature will play an important role in the classification or detection task. With respect to feature space reduction techniques like PCA and ICA, feature selection has the advantage of being a supervised method which takes into account the labels of the data.\\

Beyond reduction and selection a more principled way to optimize the feature set consists in learning features from scratch directly from the data. Concerning the audio data, it has been noticed that audio events with a semantic meaning, e.g. human voice, are often composed of a sequence of non-semantic atomic units with a stable and well defined acoustic signature. In practice, atomic units are easier to be classified but do not correspond to meaningful classes from a human-interpretative point of view. Moreover audio events are much harder to be classified due to the fact that audio units generally do not occur in a predefined order (think at the phonemes in a human voice). An interesting way to cope with this issue is to learn audio units from the data, for example by some clustering procedure, and subsequently evaluate the number of occurrences in the audio stream for each audio unit and using such number as a high-level feature. The approach is very similar to the Bag-of-Words \citep{Grauman2005} method, widely adopted in computer vision, where a codebook of visual words is learned by clustering the set of image patches extracted from the image dataset. Next, an histogram of word occurrences is calculated by increasing, for each patch, the histogram bin of the word closest to the patch itself and finally the histogram is fed to a standard classifier.

The bag-of-word approach is directly translated into bag-of-aural-word in \citep{Carletti2013} where the role of the  patches is played by audio segments, allowing to cope with complex audio sequences made of multiple distinct audio units whose order is unknown. Moreover, robustness is achieved in presence of background noise, since the classifier can learn to ignore the corresponding words. The study is focused on a limited set of classes of interest in an audio surveillance context (scream, gunshot, glass break) mixed with background noise. 
A more complex approach is devised in \citep{Kumar2012}, where also the inner temporal structure of each audio segment is taken into account. In particular, a set of audio units is modeled in an unsupervised way by a set of HMMs able to translate the audio stream into a sequence of audio units whose histogram is fed to a random forest classifier. \\

In \citep{Conte2012}, each audio segment is directly classified into a predefined set of classes by the Learning Vector Quantization (LVQ) algorithm and the degree of confidence is estimated for each classification. At the sequence level, only the frames with a high confidence are taken into account and the final label is assigned on the base of the majority voting among the high-confidence frames. Moreover, if the winning class is close to the second winning the whole sequence is classified as uncertain. Unlike the previous approaches, here each audio segment is associated to the same set of semantic concepts (the classes) under which the whole audio sequence is classified.\\

A completely unsupervised approach suited for event detection is devised in  \citep{Chin2012}. The audio sequence is transformed into a multidimensional sequence of symbols extracted by a dictionary learning procedure (either by PCA or Nonnegative Matrix Factorization). From such sequence, audio events, represented as recurring onsets of audio structures immersed in background, are detected adopting a Motif Discovery algorithm borrowed from analysis of DNA sequences.\\

Finally, in some approaches the codebook is not learned from the data but fixed a priori, typically as a set of time-frequency-scale bases. For example, in \citep{Rabaoui2009} the dictionary is composed of Gabor atoms: for each audio sequence the atom subset best approximating the signal is found by matching pursuit. Final, features are then extracted as the mean and variance of frequency and scale of the atom subset. Fixed dictionary provides better generalization properties with respect to learned dictionary at the cost of an overcomplete dictionary of higher size. 
\section{Conclusions}
In this paper, we presented an essay of automated surveillance methods based on, or including, audio devices.\\
It has two important features that make it
different and appealing with respect to the other literature reviews on similar topics. First, it proposes a global taxonomy
encompassing all the typical tasks of a surveillance systems, from the low-level ones, like background subtraction, to the semantic analysis
of a whole scene. This has never been done before for the audio sensory modality.\\
Second, it is application oriented, i.e., for each proposed method, pros and cons are discussed with respect to the needs and challenges that characterize a surveillance scenario. Moreover, though the majority of the described methods was originally proposed for surveillance and monitoring applications, the reviews includes also more general works that may likely have an impact on future developments of the field.
Beyond the general taxonomy, a set of tables and diagrams is provided to the reader in order to have quick hints concerning the best methods to be adopted for very specific tasks or operative conditions.
In conclusion, we hope that this application-oriented analysis may help in speeding up the advancement in audio-based automated surveillance, leading to the design of complete systems, able to face all the above described tasks at once, and providing at the same time convincing and robust performance.



\bibliographystyle{apalike}
\bibliography{audio_surveillance_bib,MARCObiblio}

\begin{thebibliography}{}

\bibitem[Aarabi and Zaky, 2001]{Aarabi2001}
Aarabi, P. and Zaky, S.~G. (2001).
\newblock Robust sound localization using multi-source audiovisual information
  fusion.
\newblock {\em Information Fusion}, 2(3):209--223.

\bibitem[Abu-El-Quran et~al., 2006]{Abu-El-Quran2006}
Abu-El-Quran, A., Goubran, R., and Chan, A. (2006).
\newblock Security monitoring using microphone arrays and audio classification.
\newblock {\em Instrumentation and Measurement, IEEE Transactions on},
  55(4):1025 --1032.

\bibitem[Andersson et~al., 2010]{Andersson2010}
Andersson, M., Ntalampiras, S., Ganchev, T., Rydell, J., Ahlberg, J., and
  Fakotakis, N. (2010).
\newblock Fusion of acoustic and optical sensor data for automatic fight
  detection in urban environments.
\newblock In {\em Information Fusion (FUSION), 2010 13th Conference on}, pages
  1 --8.

\bibitem[Arulampalam et~al., 2002]{Arulampalam2002}
Arulampalam, M.~S., Maskell, S., Gordon, N., and Clapp, T. (2002).
\newblock A tutorial on particle filters for online nonlinear/non-gaussian
  bayesian tracking.
\newblock {\em Signal Processing, IEEE Transactions on}, 50(2):174--188.

\bibitem[Atrey et~al., 2010]{Atrey2010}
Atrey, P.~K., Hossain, M.~A., El~Saddik, A., and Kankanhalli, M.~S. (2010).
\newblock Multimodal fusion for multimedia analysis: a survey.
\newblock {\em Multimedia Systems}, 16(6):345--379.

\bibitem[Atrey et~al., 2006a]{Atrey2006_2}
Atrey, P.~K., Kankanhalli, M.~S., and Jain, R. (2006a).
\newblock Information assimilation framework for event detection in multimedia
  surveillance systems.
\newblock {\em Multimedia Systems}, 12:239--253.

\bibitem[Atrey et~al., 2006b]{Atrey2006}
Atrey, P.~K., Maddage, N.~C., and Kankanhalli, M. (2006b).
\newblock Audio based event detection for multimedia surveillance.
\newblock In {\em Acoustics, Speech and Signal Processing, 2006. ICASSP 2006
  Proceedings. 2006 IEEE International Conference on}, volume~5, pages V --V.

\bibitem[Ayoun and Smets, 2001]{ayoun2001data}
Ayoun, A. and Smets, P. (2001).
\newblock Data association in multi-target detection using the transferable
  belief model.
\newblock {\em International Journal of Intelligent Systems},
  16(10):1167--1182.

\bibitem[Azlan et~al., 2005]{Azlan2005}
Azlan, M., Cartwright, I., Jones, N., Quirk, T., and West, G. (2005).
\newblock Multimodal monitoring of the aged in their own homes.
\newblock In {\em Proceedings of the 3rd Inteernational Conference on Smart
  Homes and Health Telematics (ICOST'05)}.

\bibitem[Barzelay and Schechner, 2010]{Barzelay2010}
Barzelay, Z. and Schechner, Y. (2010).
\newblock Onsets coincidence for cross-modal analysis.
\newblock {\em Multimedia, IEEE Transactions on}, 12(2):108--120.

\bibitem[Beal et~al., 2003]{Beal2003}
Beal, M.~J., Jojic, N., and Attias, H. (2003).
\newblock A graphical model for audiovisual object tracking.
\newblock {\em IEEE Transactions on Pattern Analysis and Machine Intelligence},
  25:828--836.

\bibitem[Blatt and Hero, 2006]{Blatt2006}
Blatt, D. and Hero, A. (2006).
\newblock Energy-based sensor network source localization via projection onto
  convex sets.
\newblock {\em Signal Processing, IEEE Transactions on}, 54(9):3614--3619.

\bibitem[Brandstein and Silverman, 1997]{Brandstein1997}
Brandstein, M.~S. and Silverman, H.~F. (1997).
\newblock A practical methodology for speech source localization with
  microphone arrays.
\newblock {\em Computer Speech \& Language}, 11(2):91 -- 126.

\bibitem[Bregman, 2005]{Bregman2005}
Bregman, A. (2005).
\newblock {\em Auditory Scene Analysis: The Perceptual Organization of Sound}.
\newblock MIT Press, London, UK.

\bibitem[Carletti et~al., 2013]{Carletti2013}
Carletti, V., Foggia, P., Percannella, G., Saggese, A., Strisciuglio, N., and
  Vento, M. (2013).
\newblock Audio surveillance using a bag of aural words classifier.
\newblock In {\em Advanced Video and Signal Based Surveillance (AVSS), 2013
  10th IEEE International Conference on}, pages 81--86.

\bibitem[Castro et~al., 2011]{Castro2011}
Castro, J., Delgado, M., Medina, J., and Ruiz-Lozano, M. (2011).
\newblock Intelligent surveillance system with integration of heterogeneous
  information for intrusion detection.
\newblock {\em Expert Systems with Applications}, 38(9):11182 -- 11192.

\bibitem[Chang et~al., 2001]{Chang2001}
Chang, S.-F., Sikora, T., and Purl, A. (2001).
\newblock Overview of the mpeg-7 standard.
\newblock {\em Circuits and Systems for Video Technology, IEEE Transactions
  on}, 11(6):688--695.

\bibitem[Chen et~al., 2013]{Bo-Wei2013}
Chen, B.-W., Chen, C.-Y., and Wang, J.-F. (2013).
\newblock Smart homecare surveillance system: Behavior identification based on
  state-transition support vector machines and sound directivity pattern
  analysis.
\newblock {\em Systems, Man, and Cybernetics: Systems, IEEE Transactions on},
  43(6):1279--1289.

\bibitem[Chen et~al., 2002]{Chen2002}
Chen, J., Hudson, R., and Yao, K. (2002).
\newblock Maximum-likelihood source localization and unknown sensor location
  estimation for wideband signals in the near-field.
\newblock {\em Signal Processing, IEEE Transactions on}, 50(8):1843--1854.

\bibitem[Chen et~al., 2005]{Chen2005}
Chen, J., Kam, A., Zhang, J., Liu, N., and Shue, L. (2005).
\newblock Bathroom activity monitoring based on sound.
\newblock In Gellersen, H., Want, R., and Schmidt, A., editors, {\em Pervasive
  Computing}, volume 3468 of {\em Lecture Notes in Computer Science}, pages
  65--76. Springer Berlin / Heidelberg.

\bibitem[Chin and Burred, 2012]{Chin2012}
Chin, M.~L. and Burred, J.~J. (2012).
\newblock Audio event detection based on layered symbolic sequence
  representations.
\newblock In {\em Acoustics, Speech and Signal Processing (ICASSP), 2012 IEEE
  International Conference on}, pages 1953--1956.

\bibitem[Choi et~al., 2005]{Choi2005}
Choi, C., Kong, D., Lee, S., Park, K., Hong, S., Lee, H., Bang, S., Lee, Y.,
  and Kim, S. (2005).
\newblock Real-time audio-visual localization of user using microphone array
  and vision camera.
\newblock In {\em Intelligent Robots and Systems, 2005. (IROS 2005). 2005
  IEEE/RSJ International Conference on}, pages 1935--1940.

\bibitem[Choi et~al., 2012]{Choi2012}
Choi, W., Rho, J., Han, D., and Ko, H. (2012).
\newblock Selective background adaptation based abnormal acoustic event
  recognition for audio surveillance.
\newblock In {\em Advanced Video and Signal-Based Surveillance (AVSS), 2012
  IEEE Ninth International Conference on}, pages 118--123.

\bibitem[Chu et~al., 2009a]{Chu2009}
Chu, S., Narayanan, S., and Kuo, C.-C. (2009a).
\newblock Environmental sound recognition with time-frequency audio features.
\newblock {\em Audio, Speech, and Language Processing, IEEE Transactions on},
  17(6):1142 --1158.

\bibitem[Chu et~al., 2009b]{Chu2009_2}
Chu, S., Narayanan, S., and Kuo, C.-C. (2009b).
\newblock A semi-supervised learning approach to online audio background
  detection.
\newblock In {\em Acoustics, Speech and Signal Processing, 2009. ICASSP 2009.
  IEEE International Conference on}, pages 1629 --1632.

\bibitem[Chua et~al., 2014]{Chua2014}
Chua, T.~W., Leman, K., and Gao, F. (2014).
\newblock Hierarchical audio-visual surveillance for passenger elevators.
\newblock In {\em MultiMedia Modeling}, pages 44--55. Springer.

\bibitem[Chung et~al., 2013]{Chung2013}
Chung, Y., Oh, S., Lee, J., Park, D., Chang, H., and Kim, S. (2013).
\newblock Automatic detection and recognition of pig wasting diseases using
  sound data in audio surveillance systems.
\newblock {\em Sensors}, 13(10):12929--12942.

\bibitem[Clavel et~al., 2005]{Clavel2005}
Clavel, C., Ehrette, T., and Richard, G. (2005).
\newblock Events detection for an audio-based surveillance system.
\newblock In {\em Multimedia and Expo, 2005. ICME 2005. IEEE International
  Conference on}, pages 1306 --1309.

\bibitem[Clavel et~al., 2008]{Clavel2008}
Clavel, C., Vasilescu, I., Devillers, L., Richard, G., and Ehrette, T. (2008).
\newblock Fear-type emotion recognition for future audio-based surveillance
  systems.
\newblock {\em Speech Commun.}, 50(6):487--503.

\bibitem[Conte et~al., 2012]{Conte2012}
Conte, D., Foggia, P., Percannella, G., Saggese, A., and Vento, M. (2012).
\newblock An ensemble of rejecting classifiers for anomaly detection of audio
  events.
\newblock In {\em Advanced Video and Signal-Based Surveillance (AVSS), 2012
  IEEE Ninth International Conference on}, pages 76--81.

\bibitem[Couvreur et~al., 2008]{Couvreur2008}
Couvreur, L., Bettens, F., Hancq, J., and Mancas, M. (2008).
\newblock Normalized auditory attention levels for automatic audio
  surveillance.

\bibitem[Cowling and Sitte, 2003]{Cowling2003}
Cowling, M. and Sitte, R. (2003).
\newblock Comparison of techniques for environmental sound recognition.
\newblock {\em Pattern Recognition Letters}, 24:2895--2907.

\bibitem[Cristani et~al., 2004a]{CristaniICPR2004}
Cristani, M., Bicego, M., and Murino, V. (2004a).
\newblock On-line adaptive background modelling for audio surveillance.
\newblock In {\em Proceedings of International Conference on Pattern
  Recognition (ICPR 2004)}, pages 399--402.

\bibitem[Cristani et~al., 2004b]{Cristani2004}
Cristani, M., Bicego, M., and Murino, V. (2004b).
\newblock On-line adaptive background modelling for audio surveillance.
\newblock In {\em Pattern Recognition, 2004. ICPR 2004. Proceedings of the 17th
  International Conference on}, volume~2, pages 399 -- 402 Vol.2.

\bibitem[Cristani et~al., 2010]{Cristani2010}
Cristani, M., Farenzena, M., Bloisi, D., and Murino, V. (2010).
\newblock Background subtraction for automated multisensor surveillance: a
  comprehensive review.
\newblock {\em EURASIP J. Adv. Signal Process}, 2010:43:1--43:24.

\bibitem[Crocco et~al., 2012]{Crocco2012}
Crocco, M., {Del Bue}, A., and Murino, V. (2012).
\newblock A bilinear approach to the position self-calibration of multiple
  sensors.
\newblock {\em IEEE Transactions on Signal Processing}, 60:660--673.

\bibitem[Crocco and Trucco, 2014]{Crocco2014}
Crocco, M. and Trucco, A. (2014).
\newblock Design of superdirective planar arrays with sparse aperiodic layouts
  for processing broadband signals via 3-d beamforming.
\newblock {\em Audio, Speech, and Language Processing, IEEE/ACM Transactions
  on}, 22(4):800--815.

\bibitem[Dai et~al., 2005]{Dai2005}
Dai, C., Zheng, Y., and Li, X. (2005).
\newblock Layered representation for pedestrian detection and tracking in
  infrared imagery.
\newblock In {\em Computer Vision and Pattern Recognition - Workshops, 2005.
  CVPR Workshops. IEEE Computer Society Conference on}, pages 13--13.

\bibitem[D'Arca et~al., 2013]{D'Arca2013}
D'Arca, E., Hughes, A., Robertson, N., and Hopgood, J. (2013).
\newblock Video tracking through occlusions by fast audio source localisation.
\newblock In {\em Image Processing (ICIP), 2013 20th IEEE International
  Conference on}, pages 2660--2664.

\bibitem[DiBiase et~al., 2001]{DiBiase2001}
DiBiase, J., Silverman, H.~F., and Brandstein, M.~S. (2001).
\newblock Robust localization in reverberant rooms.
\newblock In Brandstein, M. and Ward, D., editors, {\em Microphone Arrays},
  Digital Signal Processing, pages 157--180. Springer Berlin Heidelberg.

\bibitem[Dufaux et~al., 2000]{Dufaux2000}
Dufaux, A., Besacier, L., Ansorge, M., and Pellandini, F. (2000).
\newblock Automatic sound detection and recognition for noisy environment.

\bibitem[Ellis, 2001]{Ellis2001}
Ellis, D. P.~W. (2001).
\newblock Detecting alarm sounds.
\newblock In {\em In Proc. Workshop on Consistent and Reliable Acoustic Cues
  CRAC-2000}, pages 59--62.

\bibitem[Eronen et~al., 2006]{Eronen2006}
Eronen, A., Peltonen, V., Tuomi, J., Klapuri, A., Fagerlund, S., Sorsa, T.,
  Lorho, G., and Huopaniemi, J. (2006).
\newblock Audio-based context recognition.
\newblock {\em Audio, Speech, and Language Processing, IEEE Transactions on},
  14(1):321 -- 329.

\bibitem[Gaubitch et~al., 2013]{Gaubitch2013}
Gaubitch, N.~D., Kleijn, W.~B., and Heusdens, R. (2013).
\newblock Auto-localization in ad-hoc microphone arrays.
\newblock In {\em Acoustics, Speech and Signal Processing (ICASSP), 2013 IEEE
  International Conference on}, pages 106--110.

\bibitem[Gerlach et~al., 2012]{Gerlach2012}
Gerlach, S., Goetze, S., and Doclo, S. (2012).
\newblock 2d audio-visual localization in home environments using a particle
  filter.
\newblock In {\em Speech Communication; 10. ITG Symposium; Proceedings of},
  pages 1--4.

\bibitem[Gerosa et~al., 2007]{Gerosa2007}
Gerosa, L., Valensize, G., Tagliasacchi, M., Antonacci, F., and Sarti, A.
  (2007).
\newblock Scream and gunshot detection in noisy environments.
\newblock In {\em EUSIPCO 2007, European Signal Processing Conference}.

\bibitem[Giannakopoulos et~al., 2010]{Giannakopulos2010}
Giannakopoulos, T., Makris, A., Kosmopoulos, D., Perantonis, S., and
  Theodoridis, S. (2010).
\newblock Audio-visual fusion for detecting violent scenes in videos.
\newblock In Konstantopoulos, S., Perantonis, S., Karkaletsis, V., Spyropoulos,
  C., and Vouros, G., editors, {\em Artificial Intelligence: Theories, Models
  and Applications}, volume 6040 of {\em Lecture Notes in Computer Science},
  pages 91--100. Springer Berlin Heidelberg.

\bibitem[Gonzalez, 2007]{Gonzalez2007}
Gonzalez, R. (2007).
\newblock Enhancing video surveillance with audio events.
\newblock In {\em Digital Image Computing Techniques and Applications, 9th
  Biennial Conference of the Australian Pattern Recognition Society on}, pages
  61 --66.

\bibitem[Grauman and Darrell, 2005]{Grauman2005}
Grauman, K. and Darrell, T. (2005).
\newblock The pyramid match kernel: discriminative classification with sets of
  image features.
\newblock In {\em Computer Vision, 2005. ICCV 2005. Tenth IEEE International
  Conference on}, volume~2, pages 1458--1465 Vol. 2.

\bibitem[Guo and Li, 2003]{Guo2003}
Guo, G. and Li, S.~Z. (2003).
\newblock Content-based audio classification and retrieval by support vector
  machines.
\newblock {\em Neural Networks, IEEE Transactions on}, 14(1):209--215.

\bibitem[Hang and Hu, 2010]{Hang2010}
Hang, B. and Hu, R. (2010).
\newblock Spatial audio cues based surveillance audio attention model.
\newblock In {\em Acoustics Speech and Signal Processing (ICASSP), 2010 IEEE
  International Conference on}, pages 289--292.

\bibitem[Harma et~al., 2005]{Harma2005}
Harma, A., McKinney, M., and Skowronek, J. (2005).
\newblock Automatic surveillance of the acoustic activity in our living
  environment.
\newblock In {\em Multimedia and Expo, 2005. ICME 2005. IEEE International
  Conference on}, page 4 pp.

\bibitem[Hermansky and Morgan, 1994]{Hermansky1994}
Hermansky, H. and Morgan, N. (1994).
\newblock Rasta processing of speech.
\newblock {\em Speech and Audio Processing, IEEE Transactions on},
  2(4):578--589.

\bibitem[Hershey and Movellan, 2000]{Hershey2000}
Hershey, J. and Movellan, J. (2000).
\newblock Audio-vision: Using audio-visual synchrony to locate sounds.
\newblock In {\em Advances in Neural Information Processing Systems 12}, pages
  813--819. MIT Press.

\bibitem[Ho and Sun, 2008]{Ho2008}
Ho, K. and Sun, M. (2008).
\newblock Passive source localization using time differences of arrival and
  gain ratios of arrival.
\newblock {\em Signal Processing, IEEE Transactions on}, 56(2):464--477.

\bibitem[Hoiem et~al., 2005]{Hoiem2005}
Hoiem, D., Ke, Y., and Sukthankar, R. (2005).
\newblock Solar: sound object localization and retrieval in complex audio
  environments.
\newblock volume~5, pages v/429 -- v/432 Vol. 5.

\bibitem[Hu et~al., 2010]{Hu2010}
Hu, R., Hang, B., Ma, Y., and Dong, S. (2010).
\newblock A bottom-up audio attention model for surveillance.
\newblock In {\em Multimedia and Expo (ICME), 2010 IEEE International
  Conference on}, pages 564 --567.

\bibitem[Huang et~al., 2001]{Huang2001}
Huang, Y., Benesty, J., Elko, G., and Mersereati, R. (2001).
\newblock Real-time passive source localization: a practical linear-correction
  least-squares approach.
\newblock {\em Speech and Audio Processing, IEEE Transactions on},
  9(8):943--956.

\bibitem[Hunt et~al., 1980]{Hunt1980}
Hunt, M., Lennig, M., and Mermelstein, P. (1980).
\newblock Experiments in syllable-based recognition of continuous speech.
\newblock In {\em Acoustics, Speech, and Signal Processing, IEEE International
  Conference on ICASSP '80.}, volume~5, pages 880--883.

\bibitem[Istrate et~al., 2006]{Istrate2006}
Istrate, D., Castelli, E., Vacher, M., Besacier, L., and Serignat, J.-F.
  (2006).
\newblock Information extraction from sound for medical telemonitoring.
\newblock {\em Information Technology in Biomedicine, IEEE Transactions on},
  10(2):264 --274.

\bibitem[Ito et~al., 2009]{Ito2009}
Ito, A., Aiba, A., Ito, M., and Makino, S. (2009).
\newblock Detection of abnormal sound using multi-stage gmm for surveillance
  microphone.
\newblock In {\em Information Assurance and Security, 2009. IAS '09. Fifth
  International Conference on}, volume~1, pages 733 --736.

\bibitem[Izadinia et~al., 2013]{Izadinia2013}
Izadinia, H., Saleemi, I., and Shah, M. (2013).
\newblock Multimodal analysis for identification and segmentation of
  moving-sounding objects.
\newblock {\em Multimedia, IEEE Transactions on}, 15(2):378--390.

\bibitem[Johnson and Dudgeon, 1992]{Johnson1992}
Johnson, D.~H. and Dudgeon, D.~E. (1992).
\newblock {\em Array Signal Processing: Concepts and Techniques}.
\newblock Simon \& Schuster.

\bibitem[Kemp et~al., 2000]{Kemp2000}
Kemp, T., Schmidt, M., Westphal, M., and Waibel, A. (2000).
\newblock Strategies for automatic segmentation of audio data.
\newblock {\em Acoustics, Speech, and Signal Processing, IEEE International
  Conference on}, 3:1423--1426.

\bibitem[Kidron et~al., 2007]{Kidron2007}
Kidron, E., Schechner, Y., and Elad, M. (2007).
\newblock Cross-modal localization via sparsity.
\newblock {\em Signal Processing, IEEE Transactions on}, 55(4):1390 --1404.

\bibitem[Kim and Ko, 2011]{Kim2011}
Kim, K. and Ko, H. (2011).
\newblock Discriminative training of gmm via log-likelihood ratio for abnormal
  acoustic event classification in vehicular environment.
\newblock In {\em Proc. First ACIS/JNU Int Computers, Networks, Systems and
  Industrial Engineering (CNSI) Conf}, pages 348--352.

\bibitem[Kotus et~al., 2014]{Kotus2014}
Kotus, J., Lopatka, K., and Czyzewski, A. (2014).
\newblock Detection and localization of selected acoustic events in acoustic
  field for smart surveillance applications.
\newblock {\em Multimedia Tools and Applications}, 68(1):5--21.

\bibitem[Kotus et~al., 2013]{Kotus2013}
Kotus, J., Lopatka, K., Czy{\.z}ewski, A., and Bogdanis, G. (2013).
\newblock Audio-visual surveillance system for application in bank operating
  room.
\newblock In {\em Multimedia Communications, Services and Security}, pages
  107--120. Springer.

\bibitem[Kumar et~al., 2012]{Kumar2012}
Kumar, A., Dighe, P., Singh, R., Chaudhuri, S., and Raj, B. (2012).
\newblock Audio event detection from acoustic unit occurrence patterns.
\newblock In {\em ICASSP}, pages 489--492.

\bibitem[Kumar and Mittal, 2005]{Kumar2005}
Kumar, P. and Mittal, A. (2005).
\newblock A multimodal audio visible and infrared surveillance system (maviss).
\newblock In {\em Intelligent Sensing and Information Processing, 2005. ICISIP
  2005. Third International Conference on}, pages 151 -- 156.

\bibitem[Kushwaha et~al., 2008]{Kushwaha2008}
Kushwaha, M., Oh, S., Amundson, I., Koutsoukos, X., and Ledeczi, A. (2008).
\newblock Target tracking in heterogeneous sensor networks using audio and
  video sensor fusion.
\newblock In {\em Multisensor Fusion and Integration for Intelligent Systems,
  2008. MFI 2008. IEEE International Conference on}, pages 14--19.

\bibitem[Lecomte et~al., 2011]{Lecomte2011}
Lecomte, S., Lengelle, R., Richard, C., Capman, F., and Ravera, B. (2011).
\newblock Abnormal events detection using unsupervised one-class svm -
  application to audio surveillance and evaluation -.
\newblock In {\em Advanced Video and Signal-Based Surveillance (AVSS), 2011 8th
  IEEE International Conference on}, pages 124--129.

\bibitem[Lee et~al., 2010]{Lee2010}
Lee, K., Ellis, D. P.~W., and Loui, A.~C. (2010).
\newblock Detecting local semantic concepts in environmental sounds using
  markov model based clustering.
\newblock In {\em Proc. IEEE Int Acoustics Speech and Signal Processing
  (ICASSP) Conf}, pages 2278--2281.

\bibitem[Levy et~al., 2011]{Levy2011}
Levy, A., Gannot, S., and Habets, E. A.~P. (2011).
\newblock Multiple-hypothesis extended particle filter for acoustic source
  localization in reverberant environments.
\newblock {\em Audio, Speech, and Language Processing, IEEE Transactions on},
  19(6):1540--1555.

\bibitem[Li and Ma, 2009]{Li2009}
Li, Q. and Ma, H. (2009).
\newblock Gbed: group based event detection method for audio sensor networks.
\newblock In {\em MM '09: Proceedings of the seventeen ACM international
  conference on Multimedia}, pages 857--860, New York, NY, USA. ACM.

\bibitem[Li et~al., 2009]{Li2009_2}
Li, Q., Ma, H., and Zhao, D. (2009).
\newblock A neural network based framework for audio scene analysis in audio
  sensor networks.
\newblock In {\em PCM '09: Proceedings of the 10th Pacific Rim Conference on
  Multimedia}, pages 480--490, Berlin, Heidelberg. Springer-Verlag.

\bibitem[Li, 2000]{Li2000}
Li, S. (2000).
\newblock Content-based audio classification and retrieval using the nearest
  feature line method.
\newblock {\em Speech and Audio Processing, IEEE Transactions on},
  8(5):619--625.

\bibitem[Lin et~al., 2005]{Lin2005}
Lin, C.-C., Chen, S.-H., Truong, T.-K., and Chang, Y. (2005).
\newblock Audio classification and categorization based on wavelets and support
  vector machine.
\newblock {\em Speech and Audio Processing, IEEE Transactions on}, 13(5):644 --
  651.

\bibitem[Lu et~al., 2002]{Lu2002}
Lu, L., Zhang, H., and Jiang, H. (2002).
\newblock Content analysis for audio classification and segmentation.
\newblock {\em Speech and Audio Processing, IEEE Transactions on},
  10(7):504--516.

\bibitem[May et~al., 2011]{May2011}
May, T., van~de Par, S., and Kohlrausch, A. (2011).
\newblock A probabilistic model for robust localization based on a binaural
  auditory front-end.
\newblock {\em Audio, Speech, and Language Processing, IEEE Transactions on},
  19(1):1--13.

\bibitem[Megherbi et~al., 2005]{Megherbi2005}
Megherbi, N., Ambellouis, S., Colot, O., and Cabestaing, F. (2005).
\newblock Joint audio-video people tracking using belief theory.
\newblock In {\em Advanced Video and Signal Based Surveillance, 2005. AVSS
  2005. IEEE Conference on}, pages 135 -- 140.

\bibitem[Menegatti et~al., 2004]{Menegatti2004}
Menegatti, E., Mumolo, E., Nolich, M., and Pagello, E. (2004).
\newblock A surveillance system based on audio and video sensory agents
  cooperating with a mobile robot.
\newblock In {\em INTELLIGENT AUTONOMOUS SYSTEMS 8}, pages 335--343. IOS Press.

\bibitem[Mitrovic et~al., 2010]{Mitrovic2010}
Mitrovic, D., Zeppelzauer, M., and Breiteneder, C. (2010).
\newblock Chapter 3 - features for content-based audio retrieval.
\newblock In Zelkowitz, M.~V., editor, {\em Advances in Computers: Improving
  the Web}, volume~78 of {\em Advances in Computers}, pages 71 -- 150.
  Elsevier.

\bibitem[Monaci et~al., 2009]{Monaci2009}
Monaci, G., Vandergheynst, P., and Sommer, F. (2009).
\newblock Learning bimodal structure in audio visual data.
\newblock {\em Neural Networks, IEEE Transactions on}, 20(12):1898--1910.

\bibitem[Moncrieff et~al., 2007]{Moncrieff2007}
Moncrieff, S., Venkatesh, S., and West, G. (2007).
\newblock Online audio background determination for complex audio environments.
\newblock {\em ACM Trans. Multimedia Comput. Commun. Appl.}, 3(2):8.

\bibitem[Moragues et~al., 2011]{Moragues2011}
Moragues, J., Serrano, A., Vergara, L., and Gosalbez, J. (2011).
\newblock Improving detection of acoustic signals by means of a time and
  frequency multiple energy detector.
\newblock {\em Signal Processing Letters, IEEE}, 18(8):458 --461.

\bibitem[Ntalampiras et~al., 2009a]{Ntalampiras2009}
Ntalampiras, S., Potamitis, I., and Fakotakis, N. (2009a).
\newblock An adaptive framework for acoustic monitoring of potential hazards.
\newblock {\em EURASIP J. Audio Speech Music Process.}, 2009:2--2.

\bibitem[Ntalampiras et~al., 2009b]{Ntalampiras2009_1}
Ntalampiras, S., Potamitis, I., and Fakotakis, N. (2009b).
\newblock On acoustic surveillance of hazardous situations.
\newblock In {\em ICASSP '09: Proceedings of the 2009 IEEE International
  Conference on Acoustics, Speech and Signal Processing}, pages 165--168,
  Washington, DC, USA. IEEE Computer Society.

\bibitem[Ntalampiras et~al., 2011]{Ntalampiras2011}
Ntalampiras, S., Potamitis, I., and Fakotakis, N. (2011).
\newblock Probabilistic novelty detection for acoustic surveillance under
  real-world conditions.
\newblock {\em Multimedia, IEEE Transactions on}, 13(4):713 --719.

\bibitem[O'Donovan et~al., 2007]{O'Donovan2007}
O'Donovan, A., Duraiswami, R., and Neumann, J. (2007).
\newblock Microphone arrays as generalized cameras for integrated audio visual
  processing.
\newblock In {\em Computer Vision and Pattern Recognition, 2007. CVPR '07. IEEE
  Conference on}, pages 1 --8.

\bibitem[Peltonen et~al., 2002]{Peltonen2002}
Peltonen, V., Tuomi, J., Klapuri, A., Huopaniemi, J., and Sorsa, T. (2002).
\newblock Computational auditory scene recognition.
\newblock In {\em Acoustics, Speech, and Signal Processing, 2002. Proceedings.
  (ICASSP '02). IEEE International Conference on}, volume~2, pages 1941 --1944.

\bibitem[Pham and Cousin, 2013]{pham2013streaming}
Pham, C. and Cousin, P. (2013).
\newblock Streaming the sound of smart cities: Experimentations on the
  smartsantander test-bed.
\newblock In {\em Green Computing and Communications (GreenCom), 2013 IEEE and
  Internet of Things (iThings/CPSCom), IEEE International Conference on and
  IEEE Cyber, Physical and Social Computing}, pages 611--618. IEEE.

\bibitem[Pham et~al., 2010]{Pham2010}
Pham, Q., Lapeyronnie, A., Baudry, C., Lucat, L., Sayd, P., Ambellouis, S.,
  Sodoyer, D., Flancquart, A., Barcelo, A.-C., Heer, F., Ganansia, F., and
  Delcourt, V. (2010).
\newblock Audio-video surveillance system for public transportation.
\newblock In {\em Image Processing Theory Tools and Applications (IPTA), 2010
  2nd International Conference on}, pages 47 --53.

\bibitem[Rabaoui et~al., 2008]{Rabaoui2008}
Rabaoui, A., Davy, M., Rossignol, S., and Ellouze, N. (2008).
\newblock Using one-class svms and wavelets for audio surveillance.
\newblock {\em Information Forensics and Security, IEEE Transactions on},
  3(4):763 --775.

\bibitem[Rabaoui et~al., 2009]{Rabaoui2009}
Rabaoui, A., Lachiri, Z., and Ellouze, N. (2009).
\newblock Using hmm-based classifier adapted to background noises with improved
  sounds features for audio surveillance application.
\newblock {\em International Journal of Signal Processing}, 5(1):46--55.

\bibitem[Radhakrishnan et~al., 2005a]{Radhakrishnan2005}
Radhakrishnan, R., Divakaran, A., and Smaragdis, A. (2005a).
\newblock Audio analysis for surveillance applications.
\newblock In {\em Applications of Signal Processing to Audio and Acoustics,
  2005. IEEE Workshop on}, pages 158 -- 161.

\bibitem[Radhakrishnan et~al., 2005b]{Radhakrishnan2005_1}
Radhakrishnan, R., Divakaran, A., and Smaragdis, P. (2005b).
\newblock Systematic acquisition of audio classes for elevator surveillance.
\newblock In {\em Proc. of SPIE}.

\bibitem[Ramos et~al., 2010]{Ramos2010}
Ramos, J., Siddiqi, S., Dubrawski, A., Gordon, G., and Sharma, A. (2010).
\newblock Automatic state discovery for unstructured audio scene
  classification.
\newblock In {\em Acoustics Speech and Signal Processing (ICASSP), 2010 IEEE
  International Conference on}, pages 2154 --2157.

\bibitem[Raty, 2010]{Raty2010}
Raty, T.~D. (2010).
\newblock Survey on contemporary remote surveillance systems for public safety.
\newblock {\em Systems, Man, and Cybernetics, Part C: Applications and Reviews,
  IEEE Transactions on}, 40(5):493 --515.

\bibitem[Reynolds, 1994]{Reynolds1994}
Reynolds, D. (1994).
\newblock Experimental evaluation of features for robust speaker
  identification.
\newblock {\em Speech and Audio Processing, IEEE Transactions on},
  2(4):639--643.

\bibitem[Rouas et~al., 2006]{Rouas2006}
Rouas, J.-L., Louradour, J., and Ambellouis, S. (2006).
\newblock Audio events detection in public transport vehicle.
\newblock pages 733 --738.

\bibitem[Sasou, 2011]{Sasou2011}
Sasou, A. (2011).
\newblock Acoustic surveillance based on higher-order local auto-correlation.
\newblock In {\em Machine Learning for Signal Processing (MLSP), 2011 IEEE
  International Workshop on}, pages 1--5.

\bibitem[Scheirer and Slaney, 1997]{Scheirer1997}
Scheirer, E. and Slaney, M. (1997).
\newblock Construction and evaluation of a robust multifeature speech/music
  discriminator.
\newblock In {\em Acoustics, Speech, and Signal Processing, 1997. ICASSP-97.,
  1997 IEEE International Conference on}, volume~2, pages 1331--1334 vol.2.

\bibitem[Sheng and Hu, 2005]{Sheng2005}
Sheng, X. and Hu, Y. (2005).
\newblock Maximum likelihood multiple-source localization using acoustic energy
  measurements with wireless sensor networks.
\newblock {\em Signal Processing, IEEE Transactions on}, 53(1):44--53.

\bibitem[Smaragdis, 2003]{Smaragdis2003}
Smaragdis, P. (2003).
\newblock Audio/visual independent components.
\newblock In {\em in Proc. of International Symposium on Independant Component
  Analysis and Blind Source Separation}.

\bibitem[Smeaton and McHugh, 2005]{Smeaton2005}
Smeaton, A.~F. and McHugh, M. (2005).
\newblock Towards event detection in an audio-based sensor network.
\newblock In {\em VSSN '05: Proceedings of the third ACM international workshop
  on Video surveillance \& sensor networks}, pages 87--94, New York, NY, USA.
  ACM.

\bibitem[Souli and Lachiri, 2011]{Souli2011}
Souli, S. and Lachiri, Z. (2011).
\newblock Environmental sounds classification based on visual features.
\newblock In {\em Progress in Pattern Recognition, Image Analysis, Computer
  Vision, and Applications}, pages 459--466. Springer.

\bibitem[Strobel et~al., 2001]{Strobel2001}
Strobel, N., Spors, S., and Rabenstein, R. (2001).
\newblock Joint audio-video object localization and tracking.
\newblock {\em Signal Processing Magazine, IEEE}, 18(1):22 --31.

\bibitem[Tzanetakis and Cook, 2002]{Tzanetakis2002}
Tzanetakis, G. and Cook, P. (2002).
\newblock Musical genre classification of audio signals.
\newblock {\em Speech and Audio Processing, IEEE Transactions on},
  10(5):293--302.

\bibitem[Uzkent et~al., 2012]{Uzkent2012}
Uzkent, B., Barkana, B.~D., and Cevikalp, H. (2012).
\newblock Non-speech environmental sound classification using svms with a new
  set of features.
\newblock {\em International Journal of Innovative Computing, Information and
  Control ICIC International}.

\bibitem[Vacher et~al., 2004]{Vacher2004}
Vacher, M., Istrate, D., Besacier, L., Serignat, J., and Castelli, E. (2004).
\newblock Sound detection and classification for medical telesurvey.
\newblock In ACTA~Press, C., editor, {\em Proc. 2nd Conference on Biomedical
  Engineering}, pages 395--398, Innsbruck, Austria.

\bibitem[Valenzise et~al., 2007]{Valensize2007}
Valenzise, G., Gerosa, L., Tagliasacchi, M., Antonacci, F., and Sarti, A.
  (2007).
\newblock Scream and gunshot detection and localization for audio-surveillance
  systems.
\newblock In {\em Advanced Video and Signal Based Surveillance, 2007. AVSS
  2007. IEEE Conference on}, pages 21 --26.

\bibitem[Valera and Velastin, 2005]{valera:192}
Valera, M. and Velastin, S. (2005).
\newblock Intelligent distributed surveillance systems: a review.
\newblock {\em IEE Proceedings - Vision, Image, and Signal Processing},
  152(2):192--204.

\bibitem[Valero and Alias, 2012a]{Valero2012}
Valero, X. and Alias, F. (2012a).
\newblock Classification of audio scenes using narrow-band autocorrelation
  features.
\newblock In {\em Signal Processing Conference (EUSIPCO), 2012 Proceedings of
  the 20th European}, pages 2012--2019.

\bibitem[Valero and Alias, 2012b]{Valero2012a}
Valero, X. and Alias, F. (2012b).
\newblock Gammatone cepstral coefficients: Biologically inspired features for
  non-speech audio classification.
\newblock {\em Multimedia, IEEE Transactions on}, 14(6):1684--1689.

\bibitem[Van~Trees, 2002]{VanTrees2002}
Van~Trees, H. (2002).
\newblock {\em Detection, Estimation, and Modulation Theory, Optimum Array
  Processing}.
\newblock Detection, Estimation, and Modulation Theory. Wiley.

\bibitem[Viet et~al., 2013]{Viet2013}
Viet, Q.~N., Kang, H., Chung, S.-T., Cho, S., Lee, K., and Seol, T. (2013).
\newblock Real-time audio surveillance system for ptz camera.
\newblock In {\em Advanced Technologies for Communications (ATC), 2013
  International Conference on}, pages 392--397.

\bibitem[Vinet et~al., 2002]{Vinet2002}
Vinet, H., Herrera, P., and Pachet, F. (2002).
\newblock The cuidado project.
\newblock Content Based Retrieval.

\bibitem[Vu et~al., 2006]{Vu2006}
Vu, V.~T., Bremond, F., Davini, G., Thonnat, M., Pham, Q.-C., Allezard, N.,
  Sayd, P., Rouas, J.-L., Ambellouis, S., and Flancquart, A. (2006).
\newblock Audio-video event recognition system for public transport security.
\newblock In {\em Crime and Security, 2006. The Institution of Engineering and
  Technology Conference on}, pages 414 --419.

\bibitem[Ward et~al., 2003]{Ward2003}
Ward, D.~B., Lehmann, E., and Williamson, R. (2003).
\newblock Particle filtering algorithms for tracking an acoustic source in a
  reverberant environment.
\newblock {\em Speech and Audio Processing, IEEE Transactions on},
  11(6):826--836.

\bibitem[Weng and Guentchev, 2001]{Weng2001}
Weng, J. and Guentchev, K.~Y. (2001).
\newblock Three-dimensional sound localization from a compact non-coplanar
  array of microphones using tree-based learning.
\newblock {\em The Journal of the Acoustical Society of America},
  110(1):310--323.

\bibitem[Willert et~al., 2006]{Willert2006}
Willert, V., Eggert, J., Adamy, J., Stahl, R., and Korner, E. (2006).
\newblock A probabilistic model for binaural sound localization.
\newblock {\em Systems, Man, and Cybernetics, Part B: Cybernetics, IEEE
  Transactions on}, 36(5):982--994.

\bibitem[Xiaoling and Layuan, 2008]{Xiaoling2008}
Xiaoling, X. and Layuan, L. (2008).
\newblock Real time analysis of situation events for intelligent surveillance.
\newblock In {\em Computational Intelligence and Design, 2008. ISCID '08.
  International Symposium on}, volume~2, pages 122 --125.

\bibitem[Ye et~al., 2010]{Ye2010}
Ye, J., Kobayashi, T., and Higuchi, T. (2010).
\newblock Audio-based indoor health monitoring system using flac features.
\newblock In {\em Emerging Security Technologies (EST), 2010 International
  Conference on}, pages 90--95.

\bibitem[Youssef et~al., 2012]{Youssef2012}
Youssef, K., Argentieri, S., and Zarader, J.-L. (2012).
\newblock A binaural sound source localization method using auditive cues and
  vision.
\newblock In {\em Acoustics, Speech and Signal Processing (ICASSP), 2012 IEEE
  International Conference on}, pages 217--220.

\bibitem[Zajdel et~al., 2007]{Zajdel2007}
Zajdel, W., Krijnders, J., Andringa, T., and Gavrila, D. (2007).
\newblock Cassandra: audio-video sensor fusion for aggression detection.
\newblock In {\em Advanced Video and Signal Based Surveillance, 2007. AVSS
  2007. IEEE Conference on}, pages 200 --205.

\bibitem[Zhang et~al., 2008]{Zhang2008}
Zhang, C., Florencio, D., Ba, D., and Zhang, Z. (2008).
\newblock Maximum likelihood sound source localization and beamforming for
  directional microphone arrays in distributed meetings.
\newblock {\em Multimedia, IEEE Transactions on}, 10(3):538--548.

\bibitem[Zhang et~al., 2005]{Zhang2005}
Zhang, D., Gatica-Perez, D., Bengio, S., and McCowan, I. (2005).
\newblock Semi-supervised adapted hmms for unusual event detection.
\newblock volume~1, pages 611 -- 618 vol. 1.

\bibitem[Zhao et~al., 2010]{Zhao2010}
Zhao, D., Ma, H., and Liu, L. (2010).
\newblock Event classification for living environment surveillance using audio
  sensor networks.
\newblock In {\em Multimedia and Expo (ICME), 2010 IEEE International
  Conference on}, pages 528 --533.

\bibitem[Zhou et~al., 2008]{Zhou2008}
Zhou, X., Zhuang, X., Liu, M., Tang, H., Hasegawa-Johnson, M., and Huang, T.
  (2008).
\newblock Hmm-based acoustic event detection with adaboost feature selection.
\newblock In Stiefelhagen, R., Bowers, R., and Fiscus, J., editors, {\em
  Multimodal Technologies for Perception of Humans}, volume 4625 of {\em
  Lecture Notes in Computer Science}, pages 345--353. Springer Berlin /
  Heidelberg.

\bibitem[Zhuang et~al., 2010]{Zhuang2010}
Zhuang, X., Zhou, X., Hasegawa-Johnson, M.~A., and Huang, T.~S. (2010).
\newblock Real-world acoustic event detection.
\newblock {\em Pattern Recogn. Lett.}, 31:1543--1551.

\bibitem[Zieger et~al., 2009]{Zieger2009}
Zieger, C., Brutti, A., and Svaizer, P. (2009).
\newblock Acoustic based surveillance system for intrusion detection.
\newblock In {\em Advanced Video and Signal Based Surveillance, 2009. AVSS '09.
  Sixth IEEE International Conference on}, pages 314 --319.

\bibitem[Zieger and Omologo, 2008]{Zieger2008}
Zieger, C. and Omologo, M. (2008).
\newblock Acoustic event classification using a distributed microphone network
  with a gmm|svm combined algorithm.
\newblock In {\em Interspeech'08}.

\bibitem[Zotkin et~al., 2002]{Zotkin2002}
Zotkin, D.~N., Duraiswami, R., and Davis, L.~S. (2002).
\newblock Joint audio-visual tracking using particle filters.
\newblock {\em EURASIP J. Appl. Signal Process.}, 2002(1):1154--1164.

\end{thebibliography}



\end{document}